\documentclass[AMA,STIX1COL]{WileyNJD-v2}

\articletype{Article Type}%

\usepackage{graphics} 
\usepackage{blindtext}
\usepackage{epsfig} 
\usepackage{subfigure}
\usepackage{amsmath} 
\usepackage{amssymb}  
\usepackage{enumerate}
\newcommand{\norm}[1]{\left\lVert#1\right\rVert}
\newtheorem{thm}{Theorem}

\newtheorem{example}{Example}
\newtheorem{alg}{Algorithm}
\newtheorem{cor}{Corollary}

\newcommand{\E}{\mathrm{E}}
\renewcommand{\d}{\mathrm{d}}
\newcommand{\var}{\mathrm{Var}}

\newcommand{\tb}{\mathbf{t}}
\newcommand{\cb}{\mathbf{c}}
\newcommand{\rb}{\mathbf{r}}

\newcommand{\Vb}{\mathbf{V}}
\newcommand{\yb}{\mathbf{y}}
\newcommand{\Mk}{\mathbf{M}_k}
\newcommand{\MN}{\mathcal{M}_{N}}
\newcommand{\Mkpro}{\mathbf{M}_{k}^{\mathrm{proc}}}
\newcommand{\MNpro}{\mathcal{M}_{N}^{\mathrm{proc}}}
\newcommand{\epsilonb}{\mathbf{\epsilon}}

\newcommand{\mon}{\mathrm{mon}_n}

\definecolor{carmine}{rgb}{0.59, 0.0, 0.09}

\definecolor{frg}{rgb}{0.7, 0, 0.7}

\begin{document}
	
\title{\LARGE \bf
	Identification of Switched Autoregressive and Switched Autoregressive Exogenous Systems  from Large Noisy Data Sets \protect\thanks{This work was partially supported by National Institutes of Health (NIH) Grant R01 HL142732, National Science Foundation (NSF) Grant \#1808266 and the 
	International Bilateral Joint CNR-JST Lab COOPS.}}
	
	\author[1]{Sarah Hojjatinia}
	
	\author[1]{Constantino M. Lagoa*}
	
	\author[2]{Fabrizio Dabbene}
	
	\authormark{Sarah Hojjatinia \textsc{et al}}
%
	
\address[1]{\orgdiv{The Pennsylvania State University}, \orgname{School of Electrical Engineering and Computer Science}, \orgaddress{\state{University Park, PA}, \country{USA}}}
	
\address[2]{\orgdiv{CNR-IEIIT}, \orgname{Politecnico di Torino}, \orgaddress{\state{10129 Torino}, \country{Italy}}}
	
\corres{*Constantino Lagoa, The Pennsylvania State University, School
of Electrical Engineering and Computer Science, University Park, PA, USA. \email{lagoa@psu.edu}}
	
\abstract[Summary]{
	The paper introduces a novel methodologies for the identification of coefficients of switched autoregressive  and switched autoregressive exogenous linear models. We consider cases which  system's outputs are  contaminated by possibly large values of noise for the both case of   measurement noise in switched autoregressive models and process noise in switched autoregressive exogenous models. It is assumed that only partial information on the probability distribution of the noise is available. Given input-output data, 
	we aim at identifying switched system coefficients  and parameters of the distribution of the noise	which are compatible with the collected data.  
	We demonstrate the efficiency of the proposed approach with several academic examples. 
	The method is shown to be extremely effective in the situations where a large number of measurements is available; cases in which previous approaches based on polynomial or mixed-integer optimization  cannot be applied due to very large computational burden.
	}
	
	
%
\maketitle

\section{Introduction}

The interest in the study of hybrid systems has been persistently growing in the last years, due to their capability of describing real-world  processes in which continuous and discrete time dynamics coexist and interact. Besides classical automotive and chemical processes, emerging applications include computer vision, biological systems, and communication networks.

Moreover, hybrid systems can be used to efficiently approximate nonlinear dynamics,
with broad application, ranging from civil structures to robotics and systems biology,
that entail extracting information from high volume data streams
\cite{ozay2014convex,sznaier2014surviving}.
In the case of high dimensional data, nonlinear order reduction or  low dimensional sparse representations techniques  \cite{lin2006learning,hojjatinia2018parsimonious,saul2003think} are very effective in handling static data, but most do not exploit
dynamical information of the data.

In the literature, several results have been obtained for the analysis and control  of hybrid systems, formally characterizing important properties such as stability or reachability, and proposing different control designs \cite{lunze2009handbook}.
In parallel,  researchers rapidly realized that first-principle models may be hard to derive especially with the increase of diverse application fields. This sparked interest on the problem of identifying hybrid (switched) models starting  from experimental data; see for instance the tutorial paper \cite{paoletti2007identification}
and the survey \cite{garulli2012survey}.

It should be immediately pointed out that this identification problem is not a simple one, since the simultaneous presence of  continuous and discrete state variables gives it a combinatorial nature. The situation becomes further complicated in the presence of unknown-but-bounded noise. In this case the problem is in general NP-hard.  
Several approaches have been proposed to address this difficulty, see e.g.~\cite{lauer2011continuous}.
The paper \cite{roll2004identification}
reformulates the problem as a mixed-integer program.
These techniques proved to be very effective in situations involving relatively small noise levels or moderate dimensions,
but they do not appear to scale well, and 
their performance deteriorates as the noise level or problem size increase.

Of particular interest are recent approaches based on convex optimization: in \cite{Bako:2011:ISL:1963659.1963792}
some relaxation based on sparsity are proposed, while \cite{Ozay2015180} develops a moment based approach to identify the switched autoregressive exogenous system, and \cite{HOJJATINIA201714088} adapts it toward Markovian jump systems identification. 
These methods are surely more robust, and represent the choice of reference for medium-size problems
and medium values of noise, and have found applications in several contexts, ranging from segmentation problems arising in computer vision to biomedical systems. 

However, the methods still rely on the solution of rather large optimization problems. Even if the convex nature of these problems allows to limit the complexity growth, there are several situations for which their application becomes critical. For instance, identification problems cases that involve quite high  noise levels and/or large number of measurements. 

An enlightening example, which serves as a practical motivation for our developments, arises in healthcare applications: the availability of  \textit{activity tracking devices} allows to gather a large amount of information of the physical activity of an individual. Physical activity is a dynamic behavior, which in principle can be modeled as a dynamical system \cite{Lagoa2017}. Moreover,  its characteristics may significantly change depending on the time of the day, position, etc. This motivated the approach of modeling it as a switching system  \cite{conroy2019}.

In this paper, we focus on cases involving a very large number of sample points, possibly affected by large levels of noise. In this situation, polynomial/moments based approaches become ineffective, and different methodologies need to be devised.
The approach we propose builds upon the same premises as \cite{hojjatinia2018arx}, \cite{hojjatinia2019jump} and \cite{Ozay2015180}:
the starting point is the algebraic procedure due to Ma and Vidal \cite{Ma2005},
where it has been shown for noiseless processes, it is possible to identify the different subsystems in a switching system by recurring to a Generalized Principal Component Analysis (GPCA).
In particular,  we infer the parameters of each subsystem from the null space of a matrix $V_n(r)$ constructed from the input-output data $r$ via a nonlinear embedding (the Veronese map). 

The approach was extended to the case where process noise is present in \cite{Ozay2015180}, showing how the  entries of this matrix depend polynomially on the unknown noise terms. Then, the problem was formulated in an unknown-but-bounded setting, looking for  an admissible noise sequence rendering the matrix $V_n(r)$ rank deficient. This problem was then relaxed using polynomial optimization methods.

In this work, we follow the same line of reasoning, but then take a somewhat different route. First, we consider random noise, and we assume  that \textit{some} information on the noise is available. 
Then, instead of relaxing the problem, we exploit the availability of a large number of measurements and its ``averaged behavior.'' This allows us to devise an algorithm characterized by an extremely low complexity in terms of required operations. 
The ensuing optimization problem involves only the computation of the singular vector associated with the minimum singular value of a matrix that can be efficiently computed and whose size does not depend on the number of measurements.

 \subsection{Paper Organization}
 In Section~\ref{review}, previous results on switched system identification when no noise is present are reviewed. Section~\ref{stat} concentrates on the problem of switched system identification in the presence of measurement noise. The results are extended to the case of process noise in Section~\ref{noisy2}. Procedures for simultaneous estimation of systems parameters and noise parameters is described in Section~\ref{estim}. Several examples that illustrate the performance of the proposed approach are provided in Section~\ref{result}. Finally some concluding remarks are provided in Section~\ref{conclusion}.

\subsection{Notation} \label{sec:notation}

Given a scalar random variable $x\in\mathbb{R}$, we denote by $m_{d}$ its $d^{th}$ moment $\E[x^{d}]$, where $\E[\cdot]$ refers to expectation. The moments of $x$  may be computed according to the following integral 
\begin{equation} \label{moment}
m_{d}=\E[x^{d}]=\int_{-\infty}^{\infty}x^{d}\, f(x)\, \d x
\end{equation}
where $f(x)$ is the probability density function of $x$. Additionally, the  variance of $x$ is indicated by $\var(x)$. 

When some of the parameters $\theta$ of the distribution are not known, we use the notation $f(x|\theta)$ to denote the dependence of the probability density function on these unknown parameters. Throughout this paper, we assume that $f(x|\theta)$ is a continuous function of $\theta$. Obviously, this implies that the moments of the random variable are known continuous functions  of~$\theta$.

For example, if $x$ has a normal distribution with zero mean and we assume that the variance $\theta=\sigma^2$ is  not known then we have  
\[
f(x|\theta)=\dfrac{1}{ \sqrt{2\pi \theta}}e^{-x^{2}/2\theta}.
\]
The moments of $x$ as a function of $\theta$ are given by
\begin{equation}
m_{d}=E[x^{d}]=\begin{cases}
0 ~& \text{if}~ d~ \text{is odd}\\ 
\theta^{d/2} \, (d-1)!! ~& \text{if} ~d~ \text{is even}
\label{eq:mn}
\end{cases}
\end{equation}
where $!!$ denotes  double factorial ($n!!$ is  the product of all numbers from $n$ to 1 that have the same parity as $n$).

\section{Noiseless Switched System Identification: A Review} \label{review}

As a motivation for the approach presented in this paper, we review and slightly reformulate earlier results on an algebraic approach to the switched system identification. We refer the reader  to~\cite{VidalSoattoMaEtAl2003}
for 
details on this formulation. 
Consider a Switched AutoRegressive (SAR) system  of the form
\begin{align}
x_{k}=\sum_{j=1}^{n_{a}}{a_{j  \delta_k}} \; x_{k-j}+\sum_{j=1}^{n_{b}}{b_{j \delta_k}} \;u_{k-j}    \label{eq:C0}
\end{align}
where $x_{k} \in \mathbb{R}$  and $u_{k} \in \mathbb{R}$ are the output and input at time $k$, respectively. The variable $\delta_k \in \{1, . . . , n\}$ denotes the subsystem active at time $k$, where  $n$ is the total number of subsystems. Furthermore,  $a_{j  \delta_k}$ and $b_{j  \delta_k}$ denote unknown coefficients corresponding to mode $\delta_k$. Assume that the values of $u_k$, $k=-n_b+1, \dots, N-1$ and  $x_k$, $k=-n_a+1, \dots, N$ are available.

 As a first step towards an identification algorithm, we start by noting that equation~(\ref{eq:C0}) can be written in compact form as
\begin{align}
\tb_{\delta_{k}}^\top  \; \rb_{k}=0  \label{eq:D}
\end{align}
where we introduced the (known) regressor vector at time $k$  
\[
\rb_{k}=\left[x_{k},~ x_{k-1}, ~\cdots, ~x_{k-n_{a}}, ~u_{k-1}, ~\cdots, ~u_{k-{n_b}}\right]^\top 
\]
and the vector of (unknown) coefficients at time $k$ 
\begin{align*}
{\tb_{\delta_k}=} ~ \left[-1,~a_{1 \sigma(k)}, ~\cdots, ~a_{n_{a} \sigma(k)},~b_{1 \sigma(k)}, ~\cdots, ~b_{n_{b} \sigma(k)}\right]^\top .
\end{align*}
Hence, independently of which of the~$n$ submodels is active at time $k$, we have that the following equality should hold
\begin{align}
p_{n}(\rb_{k})= \prod_{i=1}^{n} {\tb_{i}^\top  \rb_{k}}=
 \nu_{n}(\rb_{k})^\top \cb_{n}=0,  \label{eq:E0}
\end{align}
where the
vector of parameters corresponding to the $i$-th submodel is denoted by $\tb_{i} \in \mathbb{R}^{n_{a}+n_{b}+1}$,   $\nu_{n}(\cdot)$ is  Veronese map of degree $n$ \cite{harris2013algebraic}, and $\cb_{n}$ is a vector whose entries are polynomial functions of unknown parameters $\tb_{i}$  (see \cite{vidal2005generalized} 
for explicit definition). 

The Veronese map,  also known as  polynomial embedding in machine learning,
 contains all monomials of order $n$ in lexicographical order. That is, given a vector $x\in \mathbb{R}^s$ and $n>0$, we have
\[
\nu_{n}\left(x
\right)
= \left[
\begin{array}{c}
\vdots \\
 x_{1}^{\alpha_{1}}x_{2}^{\alpha_{2}}\ldots x_s^{\alpha_ s} \\
\vdots
\end{array} 
\right], \quad \sum_{i=1}^{s} \alpha_i=n,\alpha_i\ge 0,
\]
and $\nu_{n}\left(x\right)\in\mathbb{R}^{\ell}$, with $\ell=\binom{n+s}{n}$.
Equation \eqref{eq:E0} holds for all $k$, and  these equalities  can be expressed in matrix form as follows
\begin{align}
\Vb_{n}(\rb) \, \cb_{n}=\left[
\begin{matrix}
\nu_{n}(\rb_{1})^\top ,~ \cdots,~ \nu_{n}(\rb_{N})^\top 
\end{matrix}
\right ]^\top  \cb_{n}=0   \label{eq:F}
\end{align}   
where $\rb$, without subscript, denotes the set of all regressor vectors. 
Clearly, we are able to identify  $\cb_{n}$ (and hence, under general conditions,  the system's parameters; see e.g., \cite{vidal2005generalized}) if and only if $\Vb_{n}(\rb)$ is rank deficient. In that case, the vector $\cb_{n}$ can
be found by computing the nullspace of $\Vb_{n}(\rb)$. To better clarify this procedure and fix the notation, we illustrate it in the following simple example.

\noindent
\begin{example}\label{ex1}
Consider a  system of order 1 ($n_a=n_b=1$) which switches between  two   different subsystems  ($n=2$) , that is
\begin{equation} \label{eq:ex-sub2}
\begin{aligned}
\text{subsystem 1}: ~~& x_{k}= a_{1} \, x_{k-1} +b_{1}\, u_{k-1}\\
\text{subsystem 2}:~~ & x_{k}= a_{2} \, x_{k-1} +b_{2}\, u_{k-1}\\
\end{aligned}
\end{equation}
We can rewrite the system as in equation \eqref{eq:D}. 
The regressor vector $\rb_k$ at time $k$ 
\begin{equation*} 
{\rb_{k}}=
\begin{bmatrix}
x_{k}&
x_{k-1}& 
u_{k-1} 
\end{bmatrix}^\top 
\end{equation*}
gives rise to the following Veronese vector  
\begin{equation} \label{eq:vn2}
\begin{aligned}
\nu_{n}(\rb_{k}) = 
\begin{bmatrix}
x_{k}^{2} \\
x_{k}\, x_{k-1}\\ 
x_{k} \, u_{k-1}\\
x_{k-1}^{2} \\
x_{k-1} \, u_{k-1}\\ 
u_{k-1}^{2} \\
\end{bmatrix}
\end{aligned}
\end{equation}
\newline
whose length is $\binom{n+n_{a}+n_b}{n}=\binom{2+1+1}{2}=6$. 
The corresponding coefficient vector $\cb_2$ assumes the form
\begin{align*}
\cb_{2}&= [1,  -(a_{1}+a_{2}), -(b_{1}+b_{2}),  a_{1}a_{2},   a_{1}b_{2}+b_{1}a_{2},  b_{1}b_{2}
]^\top 
\end{align*}
and its components can be observed to be polynomial functions of the parameters of the subsystems.
\end{example}
\subsection{A Reformulation of the Hybrid Decoupling Constraint } 

Note that the number of rows of the Veronese matrix $\Vb_n$ is equal to the number of measurements available for the regressor; i.e., in the notation of our paper, the number of rows is $N$. Therefore, very large data sets (large $N$) lead to computational problems that are ill conditioned or even impossible to solve. 
Hence, in this paper, we work with an equivalent condition that is more suitable for  the problem of SAR system identification from very large data sets. We now elaborate on this.

As previously mentioned, in the absence of noise, the SAR system identification is equivalent to finding a vector $\cb_n$ satisfying
\[
\cb_{n}^\top  \nu_{n}(\rb_{k})=0 ~~~~\text{ for all } k=1,2,\ldots N. 
\]
This is in turn equivalent to finding $\cb_n$ so that
\[
\frac{1}{N} \sum_{k=1}^N \cb_{n}^\top  \nu_{n}(\rb_{k}) \nu_{n}^\top (\rb_{k}) \, \cb_{n} = 0.
\]
As a result, for the noiseless case, identifying the coefficients of the submodels of switched system is equivalent to finding the singular vector $\cb_n$ associated with the minimum singular value of the matrix
\begin{equation} \label{eq:Mk}
\MN =\frac{1}{N} \sum_{k=1}^N \nu_{n}(\rb_{k}) \, \nu_{n}^\top (\rb_{k}) \doteq \frac{1}{N} \sum_{k=1}^N \Mk.
\end{equation}
Note that, by using this equivalent condition, we only need to consider square matrices of size  $\binom{n+n_{a}+n_b}{n}$. In other words, the size of this matrix \emph{does not depend} on the number of measurements. This is especially important when considering very large data sets.

\noindent
\begin{example}\label{ex2}
To illustrate the notation introduced, we revisit Example \ref{ex1}: in this case the matrix $\Mk $ has the form
\begin{equation} \label{eq:Mk_example}
	\Mk =\nu_{n}(\rb_{k}) \,\nu_{n}^\top (\rb_{k})=
	\begin{pmatrix}
	x_{k}^4 & x_{k}^3\, x_{k-1}  & x_{k}^3\, u_{k-1}  &  x_{k}^2\,x_{k-1}^2 &  x_{k}^2\,x_{k-1}\,u_{k-1}&  x_{k}^2\,u_{k-1}^2 \\
	*  &        x_{k}^2\,x_{k-1}^2   &  x_{k}^2\,x_{k-1}\,u_{k-1} &   x_{k}\,x_{k-1}^3 & x_{k}\,x_{k-1}^2\, u_{k-1}  & x_{k}\,x_{k-1}\,u_{k-1}^2\\
	*    & *   &              x_{k}^2\,u_{k-1}^2    &       x_{k}\,x_{k-1}^2\,u_{k-1}        & x_{k}\,x_{k-1}\,u_{k-1}^2   &  x_{k}\,u_{k-1}^3 \\
	* &    *    &       *       & x_{k-1}^4  & x_{k-1}^3\,u_{k-1}  &  x_{k-1}^2\,u_{k-1}^2 \\
	*    &    *    &  *    &  *     &        x_{k-1}^2\,u_{k-1}^2   &  x_{k-1}\,u_{k-1}^3\\
	*   &  * &   *   &   *   &  *  &           u_{k-1}^4
	\end{pmatrix}
	\end{equation}
and $\MN$ is just the time average of $\Mk $ above.

\end{example}
 \section{SAR System Identification in the Presence of Measurement Noise}\label{stat}
 
In this section,  we address the problem  of SAR system identification in presence of measurement noise. More precisely,
we consider SAR systems   of the form
  \begin{align}
  x_{k}=\sum_{j=1}^{n_{a}}{a_{j  \delta_k}} \; x_{k-j}+\sum_{j=1}^{n_{b}}{b_{j \delta_k}} \;u_{k-j}    \label{eq:C}
  \end{align}
    \begin{align}
  y_{k}= x_{k}+\eta_{k}     \label{eq:mno}
  \end{align}
where $y_k$ is observed output, which is assumed to be contaminated by (possibly large) noise $\eta_{k}$. As before,
 $x_{k} \in \mathbb{R}$ is the noiseless system output at time $k$ and $u_{k} \in \mathbb{R}$ is input at time $k$. Moreover, the variable $\delta_k \in \{1, . . . , n\}$ denotes the subsystem active at time $k$, where  $n$ is the total number of subsystems. 
 

  %

As a first step in the development of the proposed identification procedure, the following assumptions are made on the SAR system model and measurement noise.

\begin{assumption} \label{ass} 
Throughout this paper for SAR system identification it is assumed that:
\begin{enumerate} 
	\item[a.] Model orders $n_a$ and $n_b$ are available.
	\item[b.]  The number of subsystems $n$ is available, and each subsystem is ``visited'' infinitely often. More precisely, let $N_i(N)$ be the number of ``visits'' of subsystem $i$ up until time $N$. Then, for all $i=1,2,\ldots,n$
	\[
	\lim_{N \rightarrow \infty} \frac{N_i(N)}{N} > 0.
	\]
	\item[c.]  Noise $\eta_k$ is independent from $\eta_{l}$ for $k\neq l$, and identically distributed with probability density $f(\eta | \theta)$; where $\theta$ is a (low dimensional) vector of unknown parameters 			\item[d.] Moments of noise $m_d$ (up to order $d=4n$) are bounded.
	\item[e.]  Input sequence $u_{k}$  applied to the system is known and bounded; i.e., there exists a $L_u$ such that $|u(k)|\leq L_u$ for all $k$.
	\item[f.]  There exists a finite constant $L_x$ so that  $|x_{k}|\leq L_x$ for all $k$. 
\end{enumerate}  
\end{assumption}

We now provide a few comments on the assumptions made above. Assumption~\ref{ass}.a can be relaxed to assume only knowledge of upper bounds on $n_a$ and $n_b$. In this case, on top of the approach proposed, a search over the allowable values of $n_a$ and $n_b$ is needed to determine the values that better fit the data collected.

In the proposed procedure we rely on the use of estimates of the matrix $\MN $ described in~\eqref{eq:Mk} to determine the coefficients of the subsystems. In the case of large $N$, to be able to identify all subsystems we need Assumption~\ref{ass}.b so that each subsystem has a ``significant impact'' in the construction of $\MN $. Indeed, if the condition is not satisfied for some subsystem $i$, then  $\MN $ will not depend on it for large values of $N$.

In Assumption~\ref{ass}.c, we allow for incomplete knowledge of the measurement noise. More precisely, we assume that the overall ``form'' of the noise is known but some of its parameters will be estimated from the data. An example of this is zero mean iid Gaussian noise where the variance is not known and needs to be estimated together with the parameters of the subystems. 

Finally, Assumptions~\ref{ass}.d--f, are related to ``stability'' of the system and are needed to enforce boundedness of mean and variance of the quantities used to estimate the parameters of the subsystems and the parameters of the noise.

\subsection{Problem Statement and Preliminary Results}
To simplify the exposition to follow, we start discussing the case when  the parameters $\theta$ of the noise distribution are known and, hence,  we can compute its moments.  The more general case, where joint estimation of the parameters of the distribution of the noise is needed, is addressed in Section~\ref{estim}.

We start with the definition of the problem that we want to solve and provide some preliminary results that will allow us to develop efficient algorithms for estimation of the coefficients of the subsystems.
Consider the following problem:

\begin{problem}
Given Assumption~\ref{ass}, an input sequence $u_k$, $k=-n_b+1, \dots, N-1$ and noisy output measurements $y_k$, $k=-n_a+1, \dots, N$,
determine coefficients of the SAR model $a_{i,j}$, $i=1,2,\ldots, n_a$, $j=1,2,\ldots,n$, $b_{i,j}$, $i=1,2,\ldots, n_b$, $j=1,2,\ldots,n$. \end{problem}

As we have seen when discussing the noiseless case, the SAR system identification problem is equivalent 
%
%
to finding a vector in the null space of the matrix~${\mathcal{M}}_{N}$ defined in \eqref{eq:Mk}.
Under mild conditions, the null space of this matrix  has dimension one if and only if the data is compatible with the assumed model. 
However, if noise is present, $x_k$ is not known and, therefore,~${\mathcal{M}}_{N}$ cannot be  computed. In the remainder of this section, we make use of the available measurements as well as the a priori  information on the statistics of the noise to compute approximations of the matrix ${\mathcal{M}}_{N }\,$ and,  consequently, approximations of vectors in its null space. Let us start by establishing some properties of the entries of this matrix.

\vskip 0.1in \noindent
\textbf{On the Powers of $x_k$:} 
Since we do not have access to the values of the output $x_k$ to estimate the values of the quantities in equation~\eqref{eq:Mk}, we need to relate the powers of $x_k$ to the measurements and 
available information of the noise; i.e., its moments.
Note that $x_{k}$ is a (unknown) deterministic quantity. Therefore, for any integer $h$,
\begin{equation} 
x_{k}^{h}=E[x_{k}^{h}].
\end{equation}
Since $x_{k}=y_{k}-\eta_{k}$ we have 
\begin{equation} 
\begin{aligned}
x_{k}^{h}=E[x_{k}^{h}]=E[(y_{k}-\eta_{k})^{h}].
~~~\forall k=1, \, 2,\, \cdots, \, N.
\end{aligned}
\end{equation}

Assume now, for simplicity  the distribution of the noise is symmetric with respect to the origin. As a result, all odd moments are zero
(in particular, the noise is zero mean, i.e.\ $m_1=0$). We remark that this assumption is made only to simplify the calculations below, and that the approach can be extended to the non-symmetric case. 

We concentrate on computing the expected value of powers of $x_{k}$ recursively and in a closed form.
First, we give an example of how to compute the expected value of powers of $x_{k}$ for  powers  $h=1,2$.
For $h=1$, we have
\begin{equation}
x_{k}=E[x_{k}] =E[y_{k}-\eta_{k}]=E[y_{k}]-E[\eta_{k}]=E[y_{k}]-m_{1}=E[y_{k}],
\end{equation}
while, for $h=2$, we can write
\begin{equation} \label{16}
x_{k}^{2}=E[x_{k}^{2}] =E[(y_{k}-\eta_{k})^{2}]=E[y_{k}^{2}] - 2E[y_{k}\eta_{k}]+ E[\eta_{k}^{2}]
=E[y_{k}^{2}] - 2E[y_{k}\eta_{k}]+ m_2.
\end{equation}
We remark again that 
 the second moment of noise $E[\eta_{k}^{2}]=m_{2}$ is assumed to be known. To estimate the value of $E[y_{k}\eta_{k}]$, consider the following
\begin{equation} \label{17}
E[y_{k}\eta_{k}]=E[(x_{k}+\eta_{k})\eta_{k}]=E[x_{k}\eta_{k}]+E[\eta_{k}^{2}].
\end{equation}
The quantities $x_{k}$ and $\eta_{k}$ are mutually independent and, therefore,  $E[x_{k}\eta_{k}]=E[x_{k}]E[\eta_{k}]$, with $E[\eta_{k}]=m_{1}=0$. As a consequence, we have
\begin{equation} \label{18}
E[y_{k}\eta_{k}]=E[\eta_{k}^{2}],
\end{equation}
and finally  the value of equation \eqref{16} is
\begin{align}
E[x_{k}^{2}] & =E[y_{k}^{2}] - 2E[\eta_{k}^{2}]+ E[\eta_{k}^{2}]
=E[y_{k}^{2}] -  E[\eta_{k}^{2}] \\ & =E[y_{k}^{2}] -  m_{2}.\nonumber
\end{align}
The reasoning above can be generalized to any power of~$x_{k}$. More precisely, we have the following result, whose  proof is an immediate consequence of the derivations so far.

\vskip .1in
\begin{lemma} \label{lemxn}
	The   expected value of  the powers of $x_{k}$ satisfies
	\begin{align} 
	E[x_{k}^{h}] &=E[(y_{k}-\eta_{k})^{h}] =E[y_{k}^{h}] - \sum_{d=1}^{h} \binom{h}{d}\, E[x_{k}^{h-d}]\, E[\eta_{k}^{d}] \nonumber \\
	&=E[y_{k}^{h}] - \sum_{d=1}^{h} \binom{h}{d}\, E[x_{k}^{h-d}]\, m_{d} \nonumber \\
	&k=1, \, 2,\, \cdots, \, N. \label{eq:rec}
	\end{align} \vskip .2in
\end{lemma}

The result above provides a systematic way of relating the matrix $\Mk $ to the statistical properties of the measured output $y_k$ and of the noise~$\eta_k$. This relationship will be exploited later on to estimate $\MN $ from data.

\vskip .1in
\noindent
\begin{example}[Construction of $\Mk $]
To illustrate the use of the concepts above, we revisit again the example used in previous sections. 
Recall that, for this example, the matrix $\Mk $ has the form provided in equation~\eqref{eq:Mk_example}. 
Now, we can compute expected value of powers of $x_k$ in terms of expected value of powers of $y_k
$ and moments of measurement noise. More precisely, using Lemma~\ref{lemxn}, we obtain an equivalent expression for the matrix~$\Mk $  in \eqref{eq:Mk_example}, which is  provided in Figure~\ref{fig:Mk1}. 

\begin{figure*}[h!]
	\setlength\stripsep{\partopsep}%
%
	\newcommand\scalemath[2]{\scalebox{#1}{\mbox{\ensuremath{\displaystyle #2}}}}
	\setlength\stripsep{\partopsep}%
	%
	\begin{align*} \Mk  &=
	\scalemath{0.84}{
		\left(
		\begin{matrix}
		E[y_{k}^4]-6 \, m_{2}\,(E[y_{k}^2]-m_{2})-m_{4} & (E[y_{k}^3]-3\, m_{2} \, E[y_{k}]) \, E[y_{k-1}]  & (E[y_{k}^3]-3\, m_{2 }\, E[y_{k}])\, u_{k-1}   \\
		* &        ~(E[y_{k}^2]-m_{2})\,(E[y_{k-1}^2]-m_{2})   &  ~(E[y_{k}^2]-m_{2})\,E[y_{k-1}]\,u_{k-1} \\
		*  &  *   &              (E[y_{k}^2]-m_{2})\,u_{k-1}^2  \\
		*   & *    &   * \\
		* & *  &*  \\
		*  & * & *
		\end{matrix} 
		\right.
	} \quad \cdots \\ \\ & \hskip 1.8in \cdots \quad \scalemath{0.84}{
		\left.
		\begin{matrix}
		(E[y_{k}^2]-m_{2})\,(E[y_{k-1}^2]-m_{2}) &  (E[y_{k}^2]-m_{2})\,E[y_{k-1}]\,u_{k-1}&  (E[y_{k}^2]-m_{2})\,u_{k-1}^2\\
		(E[y_{k-1}^3]-3\, m_{2}\,E[y_{k-1}])\,E[y_{k}] & (E[y_{k-1}^2]-m_{2})\,E[y_{k}]\, u_{k-1}  & E[y_{k}]\,E[y_{k-1}]\,u_{k-1}^2 \\
		(E[y_{k-1}^2]-m_{2})\,E[y_{k}]\,u_{k-1}       & E[y_{k}]\,E[y_{k-1}]\,u_{k-1}^2   &   E[y_{k}]\,u_{k-1}^3 \\
		~~E[y_{k-1}^4]-6\,m_{2}\,(y_{k-1}^2-m_{2})-m_{4}   & ~~(E[y_{k-1}^3]-3\,m_2\,E[y_{k-1}])\,u_{k-1} &  ~~(E[y_{k-1}^2]-m_2)\,u_{k-1}^2 \\
		*   &        (E[y_{k-1}^2]-m2)\,u_{k-1}^2  &  E[y_{k-1}]\,u_{k-1}^3 \\
		* & *&           u_{k-1}^4
		\end{matrix}
		\right)}
	\end{align*}
	\caption{Example of construction of $\Mk $} \label{fig:Mk1}
\end{figure*}

\end{example}

\vskip .1in
\noindent
\textbf{On the Structure of $\Mk $:} 
We now provide one of the properties of the matrices $\Mk =\nu_{n}(\rb_{k}) \,\nu_{n}^\top (\rb_{k})$  that is central to the results to follow. If we look at the example above, we see that for given moments of the noise,  this new representation of $\Mk $ is an affine function of monomials of $y_k$ and $u_k$. This is a general result which is an immediate consequence of the reasoning described above and the fact that $y_k$ and  $y_l$ are independent random variables for $k\neq l$ and $u_k$ is a given deterministic signal.

\vskip .1in 
\begin{lemma} \label{lemma:Mk1}
	Assume that  the noise distribution and the input signal are given and fixed. Let $\mon (\cdot)$ denote a function that returns a vector with all monomials up to order $n$ of its argument.
	Then there exists an affine matrix function $M(\cdot)$ so that
	\begin{align*}
	\Mk =\nu_{n}(\rb_{k}) \,\nu_{n}^\top (\rb_{k}) &= 
	E\{M[\mon (y_k,\ldots,y_{k-n_a},u_{k-1},\ldots,u_{k-n_b})]\}\\& = M\{E[\mon (y_k,\ldots,y_{k-n_a},u_{k-1},\ldots,u_{k-n_b})]\}.
	\end{align*}
\end{lemma} \vskip .2in




\subsection{SAR Identification Algorithm}

As mentioned before, to identify the parameters of the SAR system, we need to be able to estimate the matrix~${\mathcal{M}}_{N}$ in equation~\eqref{eq:Mk}. It turns out that it can be done by using the available noisy measurements. More precisely, we have the following result.

\vskip .1in
\begin{thm} \label{thm1}
	Let $M(\cdot)$ and $\mon (\cdot)$ be the functions defined in Lemma~\ref{lemma:Mk1}. Define
	\[
	\widehat\MN  \doteq \frac{1}{N} \sum_{k=1}^N M[\mon (y_k, \ldots, y_{k-n_a},u_{k-1},\ldots,u_{k-n_b})].
	\]
	Then, as  $ N\rightarrow \infty$,
	\[
	\widehat\MN  - \MN  ~ \longrightarrow 0~~ \text{ a.s.} 
	\] \vskip .2in
\end{thm}

\noindent
\textbf{Proof:} See Appendix.
\vskip 0.1in
As a result, the empirical average computed using the noisy measurements (where expected values of monomials are replaced by the measured monomial values) converges to the desired matrix in equation~\eqref{eq:Mk}. Therefore we propose the following algorithm for identification of a SAR system.

\clearpage
\begin{alg}[SAR Identification]~\\
 Let $n_a$, $n_b$, $n$ and moments of the noise be given.
\begin{enumerate}[Step 1.] 
	\item Compute matrix
	\[
	\widehat\MN =\frac{1}{N} \sum_{k=1}^N M[\mon (y_k, \ldots, y_{k-n_a},u_{k-1},\ldots,u_{k-n_b})].
	\]
	\item Let $\cb_n$ be the singular vector associated with the minimum singular value of $\widehat\MN $.
	\item Determine the coefficients of the subsystems from the vector~$\cb_n$.
\end{enumerate}
\end{alg}

In order to perform Step 3 in Algorithm 1, we adopt 
polynomial differentiation algorithm for mixtures of hyperplanes, 
introduced by Vidal \cite[pp.~69--70]{vidal2003generalized}.
For the sake of completeness, we now review this algorithm. 
\vskip .1in 
\begin{alg}[Polynomial differentiation for mixtures of hyperplanes]~\\
Let the set of regressors $\rb$ be given and let $\cb_n$ be the vector computed by Algorithm~1.
\begin{enumerate}[Step 1.] 
	\item Define polynomial $\, p_{n}(\rb_{k})=  \cb_{n}^\top \, \nu_{n}(\rb_{k})$ 
	\item Let $Dp(\rb_{k})$ be the gradient of a polynomial $p$ at $\rb_{k}$. \\ \\
	for $\, i = n \, : \, 1$
	\[
	\yb_{i} \, = \, \text{argmin}_{\rb_{k} \, \in \, \rb, \, Dp_i(\rb_{k}) \,\neq \, 0 } ~~ \dfrac{|p_i(\rb_{k})|}{\norm{Dp_i(\rb_{k})}}
	\]
	\[
	\tb_{i} = \dfrac{Dp_i(\yb_{i})}{\norm{Dp_i(\yb_{i})}}
	\]
	
	\[p_{i-1}(\rb_{k})= \dfrac{p_{i}(\rb_{k})} {\tb_{i}^\top  \rb_{k}} \]
	\vskip .1in \noindent
	end

\item Assign point $\rb_{k}$ to subspace $S^i$ if $i = \text{argmin}_{l= 1, \cdots, n} |\tb_{l}^\top  \rb_{k}|$

\end{enumerate}
	\vskip .1in \noindent

\end{alg}

\section{switched autoregressive exogenous  system Identification} \label{noisy2}
We now show how the approach developed in the previous section can be adapted to the problem of identification of Switched AutoRegressive eXogenous (SARX) systems. 
Consider  SARX models  of the form
\begin{align}
y_{k}=\sum_{j=1}^{n_{a}}{a_{j  \delta_k}} \; y_{k-j}+\sum_{j=1}^{n_{b}}{b_{j \delta_k}} \;u_{k-j}+\epsilon_k   \label{eq:pn}
\end{align} 
where $\epsilon_{k}$ denotes process noise, $y_{k} \in \mathbb{R}$ is the output at time $k$ and $u_{k} \in \mathbb{R}$ is input at time $k$. As before, the variable $\delta_k \in \{1, . . . , n\}$ denotes the subsystem active at time $k$, where  $n$ is the total number of subsystems. Furthermore,  $a_{j  \delta_k}$ and $b_{j  \delta_k}$ denote unknown coefficients corresponding to mode $\delta_k$. 

The following assumptions are made on the above SARX system model and process noise.

\begin{assumption} \label{ass2} 
	For SARX system identification it is assumed that:
	\begin{enumerate} 
		\item  Model orders $n_a$ and $n_b$ are available.
		\item  The number of subsystems $n$ is available and each subsystem is ``visited'' infinitely often. See precise definition in Assumption~\ref{ass}.
		\item  Noise $\epsilon_k$ is independent from $\epsilon_l$ for $k\neq l$, and identically distributed with probability density $f(\epsilon | \theta)$; where $\theta$ is a (low dimensional) vector of unknown parameters.
		\item Moments of noise $m_d$ (up to order $d=4n$) are bounded.
		\item  Input sequence $u_{k}$  applied to the system is known and bounded. 
	\end{enumerate}  
\end{assumption}  

Again we assume that the order and number of subsystems  are given. If only upper bounds are available, we can search among allowable values and choose the ones better fit the data collected. As for the assumption on the system and noise, these are done do that the quantities used in the identification algorithms have bounded mean and variance.

Once more, for simplicity of exposition, in the reasoning below we assume that the distribution of the noise is known, so its moments $m_d$ are available. As mentioned before, estimation of the parameters of the distribution of the noise is addressed in Section~\ref{estim}. 

We start by noting that equation~\eqref{eq:pn} is equivalent to
\begin{align}
\tb_{\delta_k}^\top  \; \rb_{k}=0  \label{eq:D2}
\end{align}
where, for the case of ARX system with process noise,  the regressor at time $k$ takes the form
\[
\rb_{k}=[y_{k}-\epsilon_k,~ y_{k-1}, ~\cdots, ~y_{k-n_{a}}, ~u_{k-1}, ~\cdots, ~u_{k-{n_b}}]^\top 
\]
and the vector of unknown coefficients at time $k$ is
\begin{align*}
{\tb_{\delta_k}=} ~ [-1,~a_{1 \delta_k}, ~\cdots, ~a_{n_{a} \delta_k},~b_{1 \delta_k}, ~\cdots, ~b_{n_{b} \delta_k}]^\top .
\end{align*}
Hence, as before, independently of which of the~$n$ submodels is active at time $k$, we have
\begin{align}
P_{n}(\rb_{k})= \prod_{i=1}^{n} {\tb_{i}^\top  \rb_{k}}=\cb_{n}^\top  \nu_{n}(\rb_{k})=0,  \label{eq:E2}
\end{align}
where the vector of parameters corresponding to the $i$-th submodel is denoted by $\tb_{i} \in \mathbb{R}^{n_{a}+n_{b}+1}$,  and $\nu_{n,}(.)$ is  the Veronese map of degree $n$.
As before, the number of rows in the Veronese matrix $\Vb_n$, which consists of all the Veronese maps at time $k=1, 2, \cdots, N$, is equal to $N$ (the number of measurements available for the regressor) and,  therefore, a reformulation of the results  is needed to be able to address the problem of SARX identification from very large data sets.

The switched ARX system identification is equivalent to finding a vector $\cb_n$ satisfying
\[
\cb_{n}^\top  \nu_{n}(\rb_{k})=0 ~~\text{ for all } k=1,2,\ldots N. 
\]
This is in turn equivalent to finding a vector  $c_n$ so that
\[
\frac{1}{N} \sum_{k=1}^N \cb_{n}^\top  \nu_{n}(\rb_{k}) \nu_{n}^\top (\rb_{k})\cb_{n} = 0.
\]

Consequently, identifying the coefficients of the submodels of switched ARX system is equivalent to finding a singular vector $c_n$ associated with the minimum singular value of the noise dependent matrix 
\begin{equation} \label{eq:Mk2}
\MNpro =\frac{1}{N} \sum_{k=1}^N \nu_{n}(\rb_{k}) \, \nu_{n}^\top (\rb_{k}) \doteq \frac{1}{N} \sum_{k=1}^N \Mkpro 
\end{equation}
\vskip 0.1in
\noindent

The main difference between the SARX case and the SAR discussed in the previous section is the fact that the matrix $\MNpro $ is a function of the unmeasurable noise $\epsilon_k$ and cannot be directly computed.
%
Therefore, we use available information on the statistics of the noise to compute approximations of the matrix $\MNpro $,  and, consequently, approximations of vectors in its null space.
As a first step, we now relate the expected value of powers of $y_k-\epsilon_k$ to the noisy output and available information of the noise.

\vskip .1in
\begin{lemma} \label{lemxp}
	Consider output monomials of the form $\, e_k =  y_{k-1}^{h_1} \, \cdots \, y_{k-n_a}^{h_{n_a}}$, where $\sum_{i=1}^{n_a}{h_i}\, \leq 2n$,
	the   expected value of  the powers of multiplication of $y_k-\epsilon_k$ and $e_k$ satisfies
	\begin{align} 
	E[(y_k-\epsilon_k)^{h}\,e_k] &=E[y_{k}^{h}\,e_k] - \sum_{d=1}^{h} \binom{h}{d}\, E[(y_{k}-\epsilon_k)^{h-d}\,e_k]\, E[\epsilon_{k}^{d}] \nonumber \\
	&=E[y_{k}^{h}\,e_k] - \sum_{d=1}^{h} \binom{h}{d}\, E[(y_{k}-\epsilon_k)^{h-d}\,e_k]\, m_{d} \nonumber \\
	&k=1, \, 2,\, \cdots, \, N~~~~~~\&~~~ \forall i=0,1, \cdots, 2n-h. \label{eq:rec2}
	\end{align} \vskip .2in
\end{lemma}

Again, we can exploit the structure of the matrix $\Mkpro $ to determine high fidelity estimates from collected data. We start by emphasizing the following structural result 

\vskip .1in 
\begin{lemma} \label{lemma:Mk2}
	Assume that  the noise distribution and the input signal are given and fixed. Again, let~$\mon (\cdot)$ denote a function that returns a vector with all monomials up to order $n$ of its argument. Then there exists an affine function~$M_{proc}(\cdot)$ so that
	\begin{align*}
	\Mkpro  &= E\{M_{proc}[\mon (y_k,\ldots,y_{k-n_a},u_{k-1},\ldots,u_{k-n_b})]\}\\& = M_{proc}\{E[\mon (y_k,\ldots,y_{k-n_a},u_{k-1},\ldots,u_{k-n_b})]\}.
	\end{align*}
\end{lemma} \vskip .2in



\begin{example}[Construction of $\Mkpro$]
To better illustrate the proposed approach, we provide an example of how to construct the  matrix $\Mkpro $ required for SARX identification. 
To this end, consider the problem of identifying a SARX system  with $n=2$ subsystems of order $n_a=n_b=1$ of the form
\begin{equation} \label{eq:ex2}
\begin{aligned}
\text{subsystem 1}: ~~& y_{k}= a_{1} \, y_{k-1} +b_{1}\, u_{k-1} +\epsilon_k\\
\text{subsystem 2}:~~ & y_{k}= a_{2} \, y_{k-1} +b_{2}\, u_{k-1}  +\epsilon_k\\
\end{aligned}
\end{equation}
\vskip 0.05in
\noindent
where $\epsilon_k$ has a symmetric distribution.
We can rewrite the system as in equation \eqref{eq:E2}. In particular, the regressor vector $\rb_k$ at time $k$ 
\begin{equation*} 
{\rb_{k}}=
\begin{bmatrix}
y_{k}-\epsilon_k ~&
y_{k-1}~ & 
u_{k-1} 
\end{bmatrix}^\top 
\end{equation*}
gives rise to the following Veronese vector  
\begin{equation} 
\begin{aligned}
\nu_{n}(\rb_{k}) = 
\begin{bmatrix}
(y_{k}-\epsilon_k)^{2} \\
(y_{k}-\epsilon_k) \, y_{k-1}\\ 
(y_{k}-\epsilon_k) \, u_{k-1}\\
y_{k-1}^{2} \\
y_{k-1} \, u_{k-1}\\ 
u_{k-1}^{2} \\
\end{bmatrix}
\end{aligned}
\end{equation}
\newline
whose size is $l\times 1$, with $l= \binom{n+n_{a}+n_b}{n}=\binom{2+1+1}{2}=6$. 
The corresponding  vector $\cb_2$ as  a function of the parameters of the subsystems, assumes the form
\begin{align*}
\cb_{2}&= [1,  -(a_{1}+a_{2}), -(b_{1}+b_{2}),  a_{1}a_{2},   a_{1}b_{2}+b_{1}a_{2},  b_{1}b_{2}
]^\top .
\end{align*}
From $\rb_k$ and $\nu_{n}(\rb_k)$, we can  compute matrix $\Mkpro \,$ as follows\\

	$
	\Mkpro =\nu_{n}(\rb_k) \,\nu_{n}^\top (\rb_k)=$ \[
	\begin{pmatrix}
	(y_{k}-\epsilon_k)^4 & (y_{k}-\epsilon_k)^3\, y_{k-1}  & (y_{k}-\epsilon_k)^3\, u_{k-1}  &  (y_{k}-\epsilon_k)^2\,y_{k-1}^2 &  (y_{k}-\epsilon_k)^2\,y_{k-1}\,u_{k-1}&  (y_{k}-\epsilon_k)^2\,u_{k-1}^2 \\
	*  &        (y_{k}-\epsilon_k)^2\,y_{k-1}^2   &  (y_{k}-\epsilon_k)^2\,y_{k-1}\,u_{k-1} &   (y_{k}-\epsilon_k)\,y_{k-1}^3 & (y_{k}-\epsilon_k)\,y_{k-1}^2\, u_{k-1}  & (y_{k}-\epsilon_k)\,y_{k-1}\,u_{k-1}^2\\
	*    & *   &              (y_{k}-\epsilon_k)^2\,u_{k-1}^2    &       (y_{k}-\epsilon_k)\,y_{k-1}^2\,u_{k-1}        & (y_{k}-\epsilon_k)\,y_{k-1}\,u_{k-1}^2   &  (y_{k}-\epsilon_k)\,u_{k-1}^3 \\
	* &    *    &       *       & y_{k-1}^4  & y_{k-1}^3\,u_{k-1}  &  y_{k-1}^2\,u_{k-1}^2 \\
	*    &    *    &  *    &  *     &        y_{k-1}^2\,u_{k-1}^2   &  y_{k-1}\,u_{k-1}^3\\
	*   &  * &   *   &   *   &  *  &           u_{k-1}^4
	\end{pmatrix}.
	\]
Then, as we have the values of noisy output $y_k$, we compute expected value of powers of $y_k-\epsilon_k$ in terms of expected value of powers of $y_k
$ and moments of process noise. Following the results of Lemma \ref{lemxp}, we obtain the second matrix in Figure~\ref{fig:Mk2}. 
For system of equation \eqref{eq:ex2}, $\Mkpro $ is given by the two expression in Figure~\ref{fig:Mk2}.


\begin{figure*}[t]
	\setlength\stripsep{\partopsep}%

	%
	\newcommand\scalemath[2]{\scalebox{#1}{\mbox{\ensuremath{\displaystyle #2}}}}
	\setlength\stripsep{\partopsep}%
	%
	\begin{align*} \Mkpro  &=
	\scalemath{0.84}{
		\left(
		\begin{matrix}
		E[y_{k}^4]-6 \, m_{2}\,(E[y_{k}^2]-m_{2})-m_{4}~~ & E[y_{k}^3\,y_{k-1}]-3\, m_{2} \, E[y_{k}\,y_{k-1}]  & (E[y_{k}^3]-3\, m_{2 }\, E[y_{k}])\, u_{k-1}   \\
		* &        ~E[y_{k}^2\,y_{k-1}^2]-m_{2}\,E[y_{k-1}^2]  &  ~(E[y_{k}^2\,y_{k-1}]-m_{2}\,E[y_{k-1}])\,u_{k-1} \\
		*  &  *   &              (E[y_{k}^2]-m_{2})\,u_{k-1}^2  \\
		*   & *    &   * \\
		* & *  &*  \\
		*  & * & *
		\end{matrix} 
		\right.
	} \quad \cdots \\ \\ & \hskip 1.8in \cdots \quad \scalemath{0.84}{
	\left.
	\begin{matrix}
	E[y_{k}^2\,y_{k-1}^2]-m_{2}\,E[y_{k-1}^2]~~&  (E[y_{k}^2\,y_{k-1}]-m_{2}\,E[y_{k-1}])\,u_{k-1}&  (E[y_{k}^2]-m_{2})\,u_{k-1}^2\\
	E[y_{k}\,y_{k-1}^3] & E[y_{k}\,y_{k-1}^2]\, u_{k-1}  & E[y_{k}\,y_{k-1}]\,u_{k-1}^2 \\
	E[y_{k}\,y_{k-1}^2]\,u_{k-1}       & E[y_{k}\,y_{k-1}]\,u_{k-1}^2   &   E[y_{k}]\,u_{k-1}^3 \\
	~~E[y_{k-1}^4]   & ~~E[y_{k-1}^3]\,u_{k-1} &  ~~E[y_{k-1}^2]\,u_{k-1}^2 \\
	*   &        E[y_{k-1}^2]\,u_{k-1}^2  &  E[y_{k-1}]\,u_{k-1}^3 \\
	* & *&           u_{k-1}^4
	\end{matrix}
	\right)}
\end{align*}
\caption{Example of construction of $\Mkpro $} \label{fig:Mk2}
\end{figure*}
\end{example}
\subsection{SARX Identification Algorithm}
As mentioned before, to identify the parameters of the SARX system, we need to be able to estimate the matrix~$\MNpro $ in equation \eqref{eq:Mk2}. It turns out that it can be done by exploiting its structure and using the available noisy measurements. More precisely, we have the following result.


\begin{thm} \label{thm11}
Let $M_{proc}(\cdot)$ and $\mon (\cdot)$ be the functions defined in Lemma~\ref{lemma:Mk2}. Define
	\[
	\widehat\MNpro  \doteq \frac{1}{N} \sum_{k=1}^N M_{proc}[\mon (y_k,\ldots,y_{k-n_a},u_{k-1},\ldots,u_{k-n_b})].
	\]

\noindent	
Take any monomial 
\begin{equation} \label{mon}
\begin{aligned}
z_k = y_k^{h_0}\, y_{k-1}^{h_1} \, \cdots \, y_{k-n_a}^{h_{n_a}}\\
\end{aligned}
\end{equation}
 \vskip 0.1in \noindent 
where 
$\sum_{i=0}^{n_a}{h_i} \, \leq \, 2n\, , ~~~ h_i = 0, 1, 2, \cdots, \, 2n\, . $
If for any $h_1, h_2, \cdots, h_{n_a}$, 
the sequence $\{z_k, \, k\, \geq \, 1 
\}$ 
satisfies  
\vskip 0.05in
\begin{itemize}
    \item $
\sum_{k=1}^{\infty}{\dfrac{(\var\,z_k)(\log \, k)^2}{k^2}} ~ < ~ \infty$
\item $\sum_{l=1}^{\infty}{\dfrac{\rho_l}{l^q}} ~ < ~ \infty ~~~~~~~~ \text{for some} ~~0 \leq q < 1$ 
\end{itemize}
\vskip 0.1in \noindent
where $\{\rho_l~, l \geq 1\}$ is a sequence of constants  such that
$
\sup_{k \geq 1}{|\mathrm{Cov} \,(z_k, \, z_{k+l})|} \leq \rho_l~~~~ l \geq 1.
$
%
%
\vskip 0.1in \noindent
	then, 
	as  $ N\rightarrow \infty$,
	\[
	\widehat\MNpro - \MNpro  \longrightarrow ~0~~ \text{ a.s.} 
	\] \vskip .1in \noindent
\end{thm}

\noindent
\textbf{Proof:} Direct application of the results in~\cite{hu2008convergence} with $b_n=n$. For completeness this result is stated as Theorem~\ref{thmcov} in Appendix. 
\vskip 0.1in
The conditions of the theorem above are rather general and state that, if the output of the system is ``well-behaved'' then empirical averages of functions of the collected data can be used of estimate the matrix $\MN $ and, hence, the coefficients of the subsystems. 

Although these condition are rather abstract, it turns out that there is a an important special case 
where Theorem \ref{thm11} can be applied, namely the case when the SARX system is uniformly exponentially stable and the noise is Normally distributed.


\begin{cor} \label{thmwork}
Let the SARX system in \eqref{eq:pn} be uniformly exponentially stable, and the 
noise distribution is zero mean Normal i.e. $\epsilon_k \sim N(0,\sigma^2)$. Assume moreover that the 
dynamics of switching $\delta_k$ at time $k$ are independent from input $u_k$ and output $y_k$.
Then 
the conditions  of Theorem \ref{thm11} are satisfied, and 	therefore as  $ N\rightarrow \infty$,
\begin{equation*}
	\widehat\MNpro - \MNpro  \longrightarrow ~0~~ \text{ a.s.} 
	\end{equation*}
\end{cor}


\noindent
\textbf{Proof:} See Appendix. 
\vskip .1in

As a result, the empirical average computed using the noisy measurements (where expected values of monomials are replaced by the averages of the measured monomial values) converges to the desired matrix in equation~\eqref{eq:Mk2}. Therefore we propose the following algorithm for identification of a SARX system.

\vskip .1in 
\begin{alg}[SARX system identification]~\\
Let $n_a$, $n_b$, $n$ and some parameters of the noise be given. 
\begin{enumerate}[Step 1.] 
	\item Compute matrix
	\[
	\widehat\MNpro =\frac{1}{N} \sum_{k=1}^N M_{proc}[\mon (y_k,\ldots,y_{k-n_a},u_{k-1},\ldots,u_{k-n_b})]
	\]
	\item Let $\cb_n$ be the singular vector associated with the minimum singular value of $\widehat\MNpro $.
	\item Determine the coefficients of the subsystems from the vector~$\cb_n$.
\end{enumerate}
\end{alg}


\section{Estimating Unknown Noise Parameters } \label{estim}

We now address the case where the distribution of the noise is not completely known. In particular, as previously mentioned, in this paper it is assumed that the distribution of the noise is known except for a few parameters. 
For simplicity of exposition, let us consider the case where the noise has one scalar unknown parameter $\theta$. The reasoning can be extended to any case where the set of allowable parameters can be efficiently gridded.

In such a case, the objective is to simultaneously estimate  system parameters 
and the parameter $\theta$. We start by noting  that computing 
$\MN $ ($\MNpro $) using the 
true value of $\theta$ results in a rank deficient matrix. Moreover, given 
collected data $y_k$ and $u_k$, the matrix 
$\widehat\MN $ ($\widehat\MNpro $)
is a continuous function of the moments of  noise and, hence, a \emph{known} continuous function of the parameter $\theta$. Given previous convergence results,  the true value of $\theta$ will make  $\widehat\MN $ ($\widehat\MNpro $) to have a very small minimum singular value (especially for large values of $N$). For this reason,  estimation of $\theta$ can be performed by minimizing the minimum singular value of matrix above over the allowable values of $\theta$. More precisely, we propose the following algorithm

\vskip .1in 
\begin{alg}[Joint SARX system and noise parameter  identification]~\\
Let $n_a$, $n_b$, $n$, some parameters of the noise  and $\theta_{\max}$  be given. 
\begin{enumerate}[Step 1.] 
	\item Compute matrix $\MN $ ($\MNpro $)
	as a function of the noise parameter $\theta$.
	\item  Find the value  $\theta^* \in [0,\theta_{\max}]$ that minimizes the minimum singular value of $\widehat\MN $ ($\widehat\MNpro $).
	\item Let $\cb_n$ be associated singular vector.
	\item Determine the coefficients of the subsystems from the vector~$\cb_n$.
\end{enumerate}
\end{alg}
\vskip .1in 

Note that the optimization in Step 2 is in general nonconvex, but it can be solved via an easily implementable line-search. However, the solution~$\theta^*$ might not be unique; i.e.,  there might exist several values of $\theta$ that lead to a minimum singular value very close to zero. 
In practice, our experience has been that, for sufficiently large $N$, the above algorithm provides both a good estimate of the systems coefficients, and noise parameters; especially if we take $\theta^*$ to be the smallest value of $\theta$ for which the minimum singular value of $\widehat\MN $ ($\widehat\MNpro $) is below a given threshold.

\clearpage

\begin{table*}[ht]
	\caption{Identifying polynomial coefficients for different values of noise variance and different \textbf{SAR} system run.}
	\label{tab:table1}
	\centering
	\begin{tabular}{c|c|c|c|c|c|c||c|c|c} 
		\hline \hline
		\textbf{Experiment} &\textbf{Value 1} & \textbf{Value 2} & \textbf{Value 3}& \textbf{Value 4} & \textbf{Value 5} & \textbf{Value 6}& \textbf{Value 7}& \textbf{Value 8}& \textbf{Value 9}\\
		\text{\#} &1 & $-(a_{1}+a_{2})$ & $-(b_{1}+b_{2})$ & $a_{1}\,a_{2}$ & $a_{1}\,b_{2}+b_{1}\,a_{2}$ &$b_{1}\,b_{2}$& \text{$\gamma$}& \text{$\sigma^2$}& $\widehat{\sigma}^2$\\
		\hline \hline
		\text{true parameters}	&1 & 0.2 &0 & -0.15 & -0.8 & -1 & - & -&-\\
		\hline\hline
		\text{identification 1}	&1	&  0.2002 &  0.0001 &  -0.1503 &  -0.7989 &-0.9996& 0.2410 &0.1 & 0.1000\\
		\hline
		\text{identification 2}	&1	&  0.2002 &   0.0011 &  -0.1510 &  -0.7974&   -1.0004& 0.5187 &0.5 & 0.4980\\
		\hline
		\text{identification 3}	&1 &  0.1977  &  0.0046   &-0.1548   &-0.7997  & -0.9966& 0.6494 & 1& 1.0010\\
		\hline
		\text{identification 4}	&1 &  0.2120 &  0.0003 &  -0.1485  & -0.8006  & -1.0017  & 0.8516 & 2& 1.9950\\
		\hline \hline
	\end{tabular}
\end{table*}
\section{Numerical Results}\label{result}

In this section we present some numerical examples which illustrate the effectiveness of the proposed approach.

\subsection{SAR system identification} \label{result1}
In the following example,  we address the problem of identifying a two-mode switched system of the form \eqref{eq:C}--\eqref{eq:mno}, whose true coefficients are   $a_{1}=0.3,~ b_{1}=1,~ a_{2}=-0.5,$ and $b_{2}=-1$. Measurement noise is assumed to be zero-mean with Normal distribution.
In the numerical examples presented, $N=10^{6}$ input-output data is given. 
 True and identified coefficients for different  variances of noise, are presented in Table~\ref{tab:table1}. Variance of  noise and  noise to output ratio for each experiment are also  shown in this table. 
The provided noise to output ratio ($\gamma$) is defined as
\begin{equation}
\gamma = \dfrac{\max \, |\eta|}{\max \, |y|}.
\end{equation}

Results are as expected even for high values of noise in comparison to output. As it is illustrated in Table~\ref{tab:table1}, the identified parameters are very close to true values which   demonstrates  the convergence of  proposed algorithm even for small signal to noise ratio.
Moreover, the algorithm requires a very small computational effort. For the case of~$10^6$ measurements and using an off-the-shelf core i5 laptop with 8 Gigs of RAM, the running time  is between 7 to 8 seconds, which shows the effectiveness of approach for very large data sets.

\begin{figure}[h!]
	\begin{center}
		\centering\includegraphics[clip,width=0.5\columnwidth]{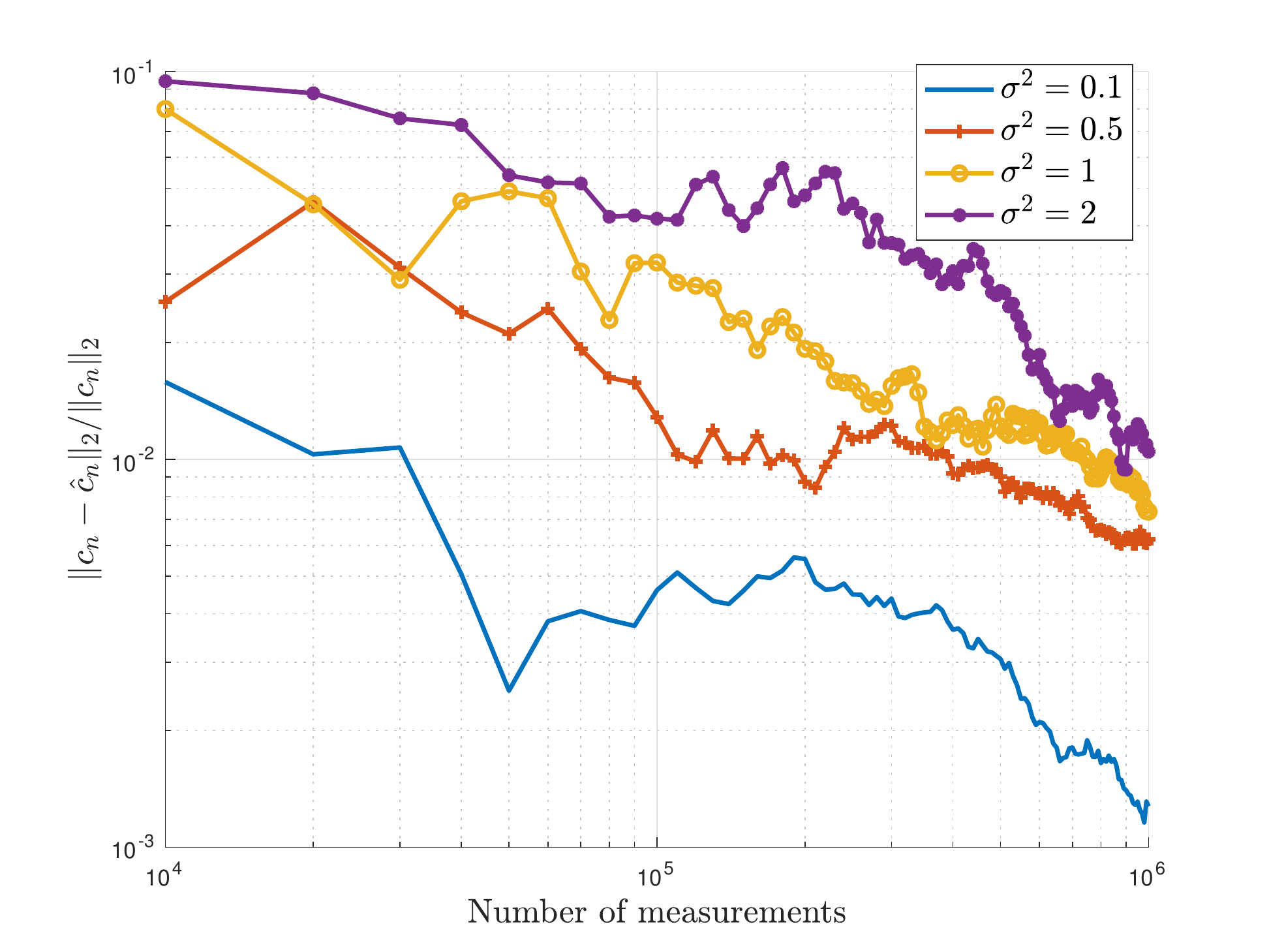}
		\caption{Estimation error of system coefficients}\label{fig:comp}
	\end{center}
\end{figure}

The error between true coefficients of system and estimated coefficients, $\|\cb_n-\hat{\cb}_n\|_2/||\cb_n||_2$, as a function of number of measurements, $N$, is depicted in Figure \ref{fig:comp} for different values of noise variance. As it can be seen from Figure \ref{fig:comp}, the error decreases as the number of measurements increases. Rate of convergence is fast, despite the fact that, in some of the experiments,  a large amount of noise is used. 
It should be noted that these results are for one experiment, and given that this is a realization of a random process, error is not always decreasing.
 For all values of noise variance, error will eventually decrease and the estimated values of coefficients converge to the true values.

Now we consider  estimation of the individual subsystems. For the above mentioned example, Table \ref{tab:table2} shows the values of subsystems coefficients for different experiments related to different  values of noise variance. As we see in this table the value of coefficients are very close to the true values, even when the noise variance is high with noise magnitude in average around  85\% of the signal magnitude.

\begin{table}[ht]
	\caption{Identifying submodels' coefficients for different values of noise variance in \textbf{SAR} systems.}
	\label{tab:table2}
	\centering
	\begin{tabular}{c|c|c|c|c|c} 
		\hline \hline
		\text{submodels' } &true&  variance & variance &variance  &variance \\
		\text{coefficients} & \text{values}& $\sigma^2=0.1$& $\sigma^2=0.5$& $\sigma^2=1$& $\sigma^2=2$\\
		\hline \hline
		$a_{1}$	&   0.3  & 0.3002 &0.2981     &0.3006      &  0.2938    \\
		\hline
		 $b_{1}$	&1	&  0.9988     &  1.0007     & 0.9412      &   1.0031    \\
		\hline
		$a_{2}$	&-0.5	&  -0.4996 &   -0.5000 &  -0.5006 &  -0.5059\\
		\hline
		$b_{2}$	&-1 &  -0.9999  &  -0.9991  & -1.0004&-1.0011 \\
		\hline \hline
	\end{tabular}
\end{table}

The estimation of noise variance based on the structure of matrix $\Mk $ is shown in Table \ref{tab:table1} as $\widehat{\sigma}^2$. The estimates of noise variance are very close to the true values of variance. By knowing the structure of matrix $\Mk \,$, the dependence of every entry on the  moments of noise, and the relation in between these moments and the unknown variance (see Section~\ref{sec:notation}), we are able to estimate the noise parameter (in this case, noise variance). This illustrates the capability of the proposed algorithm to estimate both system and noise parameters even for large values of noise.

\begin{figure}[h!]
\begin{center}
	\subfigure[Case 1: true noise variance $\sigma^2= 1$.]{
		\centering\includegraphics[trim={1cm 0.2cm 1cm 1cm},clip,width=0.4\columnwidth]{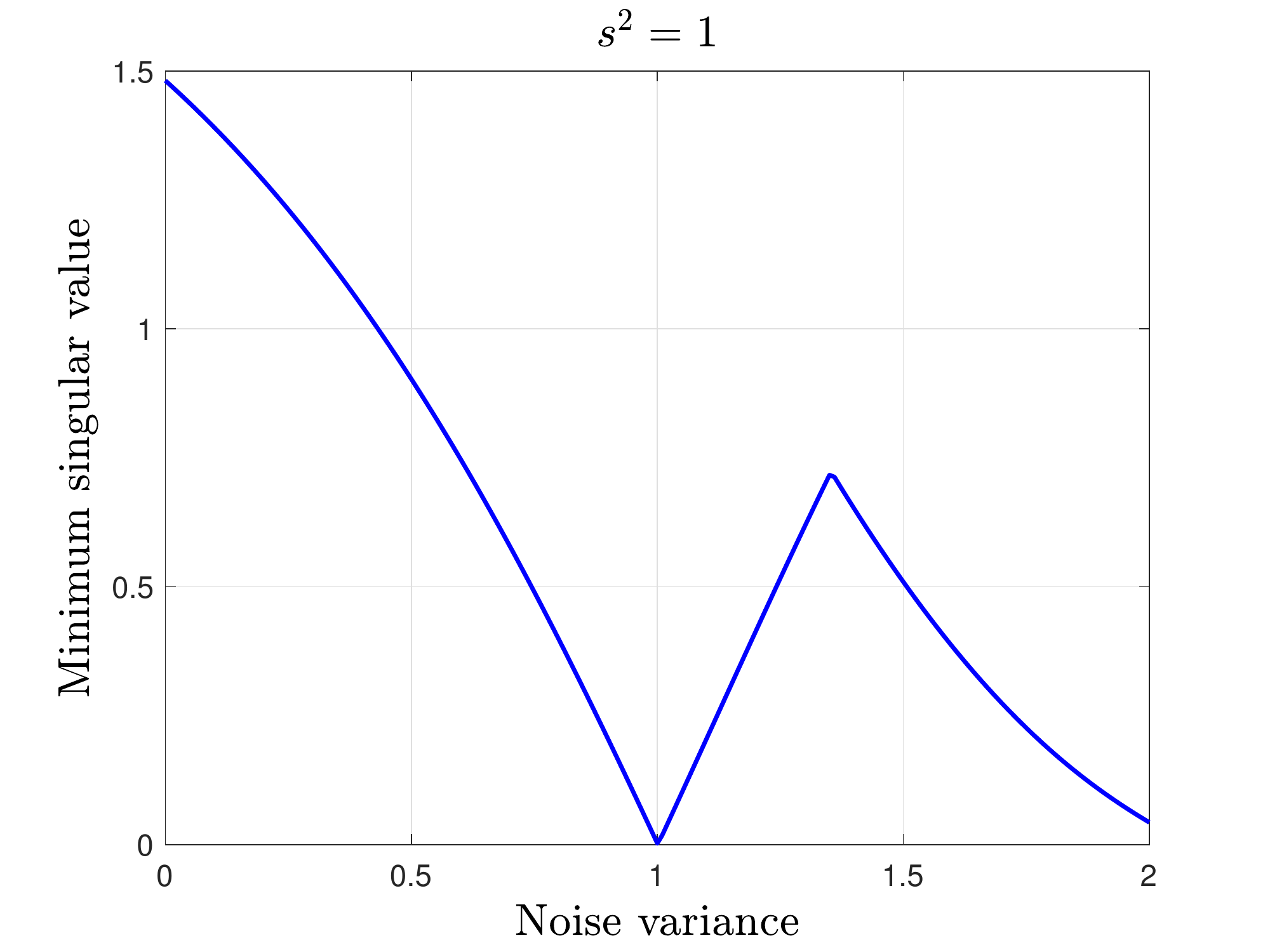} \label{fig:2a}
	}
	\subfigure[Case 2: true noise variance = $\sigma^2=2$.]{
		\centering\includegraphics[trim={1cm 0.3cm 1cm 1cm},clip,width=0.4\columnwidth]{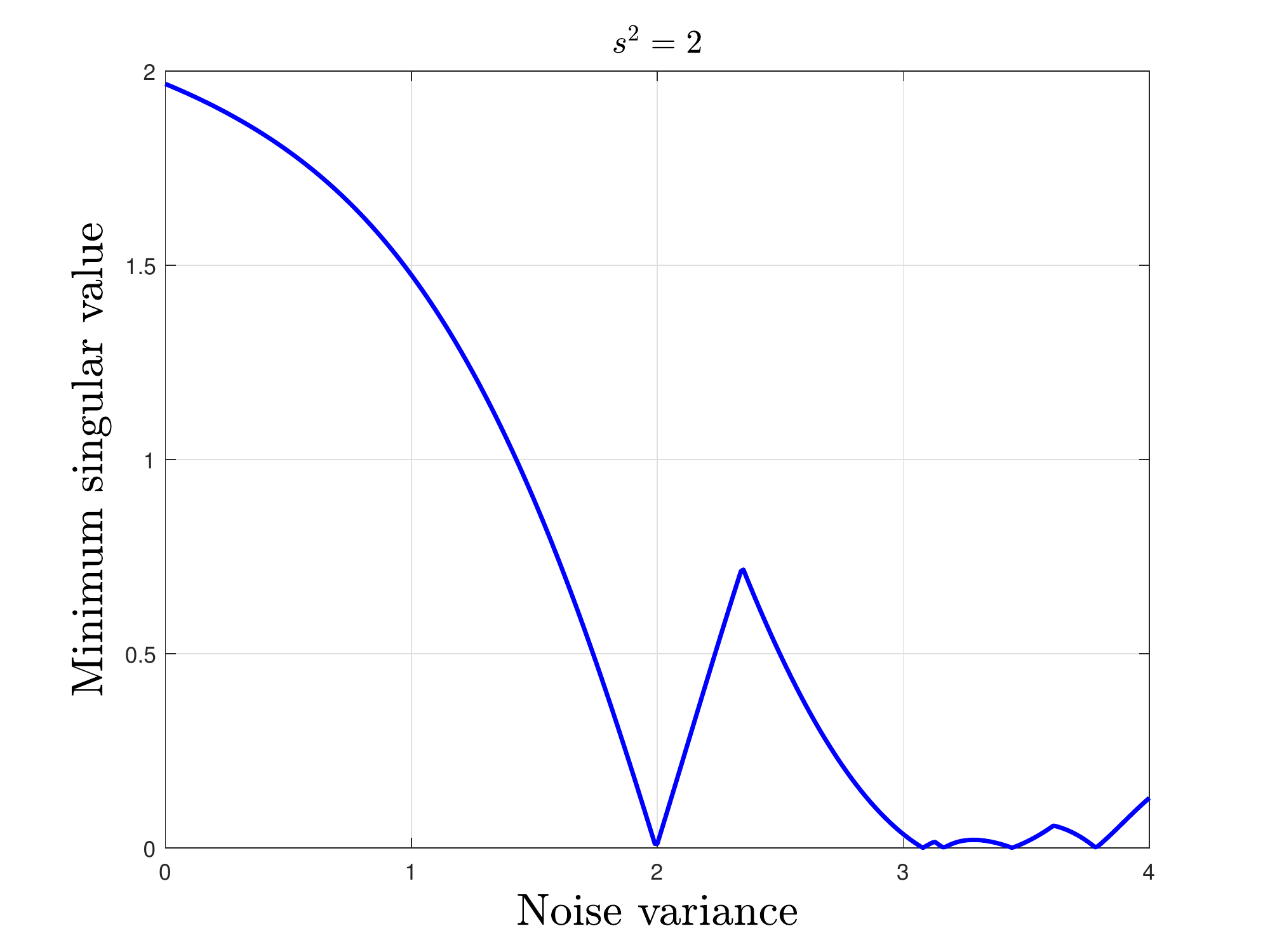} \label{fig:2b}
	}
	\caption{Estimation of noise variance using Algorithm~3}\label{fig:2}
\end{center}
\end{figure}

 Two examples of the process of estimating the unknown variance of noise are shown in Fig.~\ref{fig:2}; where Fig.~\ref{fig:2a} is for the case of given data contaminated with  noise of variance 1, and Fig.~\ref{fig:2b} is for data with measurement noise of variance 2. By taking $\sigma^*$ as the smallest local minimum, the estimated variance for both cases in Fig.~\ref{fig:2a} and Fig.~\ref{fig:2b} is very close to the true values.

\subsection{SARX system identification} \label{result2}
\begin{table*}[ht]
	\caption{Identifying polynomial coefficients for different values of noise variance and different \textbf{SARX} system run.}
	\label{tab:table3}
	\centering
	\begin{tabular}{c|c|c|c|c|c|c||c|c|c} 
		\hline \hline
		\textbf{Experiment} &\textbf{Value 1} & \textbf{Value 2} & \textbf{Value 3}& \textbf{Value 4} & \textbf{Value 5} & \textbf{Value 6}& \textbf{Value 7}& \textbf{Value 8}& \textbf{Value 9}\\
		\text{\#} &1 & $-(a_{1}+a_{2})$ & $-(b_{1}+b_{2})$ & $a_{1}\,a_{2}$ & $a_{1}\,b_{2}+b_{1}\,a_{2}$ &$b_{1}\,b_{2}$& \text{$\gamma$}& \text{$\sigma^2$}& $\widehat{\sigma}^2$\\
		\hline \hline
		\text{true parameters}	&1 & 0.2 &0 & -0.15 & -0.8 & -1 & - & -&-\\
		\hline\hline
		\text{identification 1}	&1	&  0.2006 &  -0.0011 &  -0.1504 &  -0.8011 &-1.0001& 0.2657 &0.1 & 0.1\\
		\hline 
		\text{identification 2}	&1	& 0.1991  & -0.0001&  -0.1506 &-0.8029  & -1.0007 &  0.5044   &0.5 & 0.5 \\
		\hline    
		\text{identification 3}	&1 & 0.1960&  -0.0008  &  -0.1499   &-0.7963  &-1.0032  & 0.5656  & 1& 1\\
		\hline
		\text{identification 4}	&1 & 0.2052  & 0.0084  & -0.1493  & -0.8050 & -1.0036  & 0.7649 & 2& 2\\
		\hline \hline    
	\end{tabular}
\end{table*}
In this section's examples,  we address the problem of identifying a two-mode switched system of the form of equation \eqref{eq:ex2}, whose true coefficients are, again,   $a_{1}=0.3,~ b_{1}=1,~ a_{2}=-0.5,$ and $b_{2}=-1$. Process noise is assumed to be zero-mean with Normal distribution.
A total number of $N=10^{6}$ input-output data is given for each experiment. 
True and identified coefficients for different  variances of noise, are presented in Table~\ref{tab:table3}. Noise to output ratio and estimate of noise variance for each experiment are also  shown in this table.

Once again we see that the proposed approach is very effective. As depicted  in Table~\ref{tab:table3}, the error in the identification of the system's parameters is very small which   demonstrates  the convergence of  proposed algorithm even for small signal to noise ratio.
Again, the algorithm requires a very small computational effort. For the case of~$10^6$ measurements and using the same off-shelf computer as before, the running time  is between 2 to 9 seconds. Again this shows how well the proposed approach scales with the number of measurements.

\begin{figure}[h!]
	\begin{center}
				\centering\includegraphics[clip,width=0.5\columnwidth]{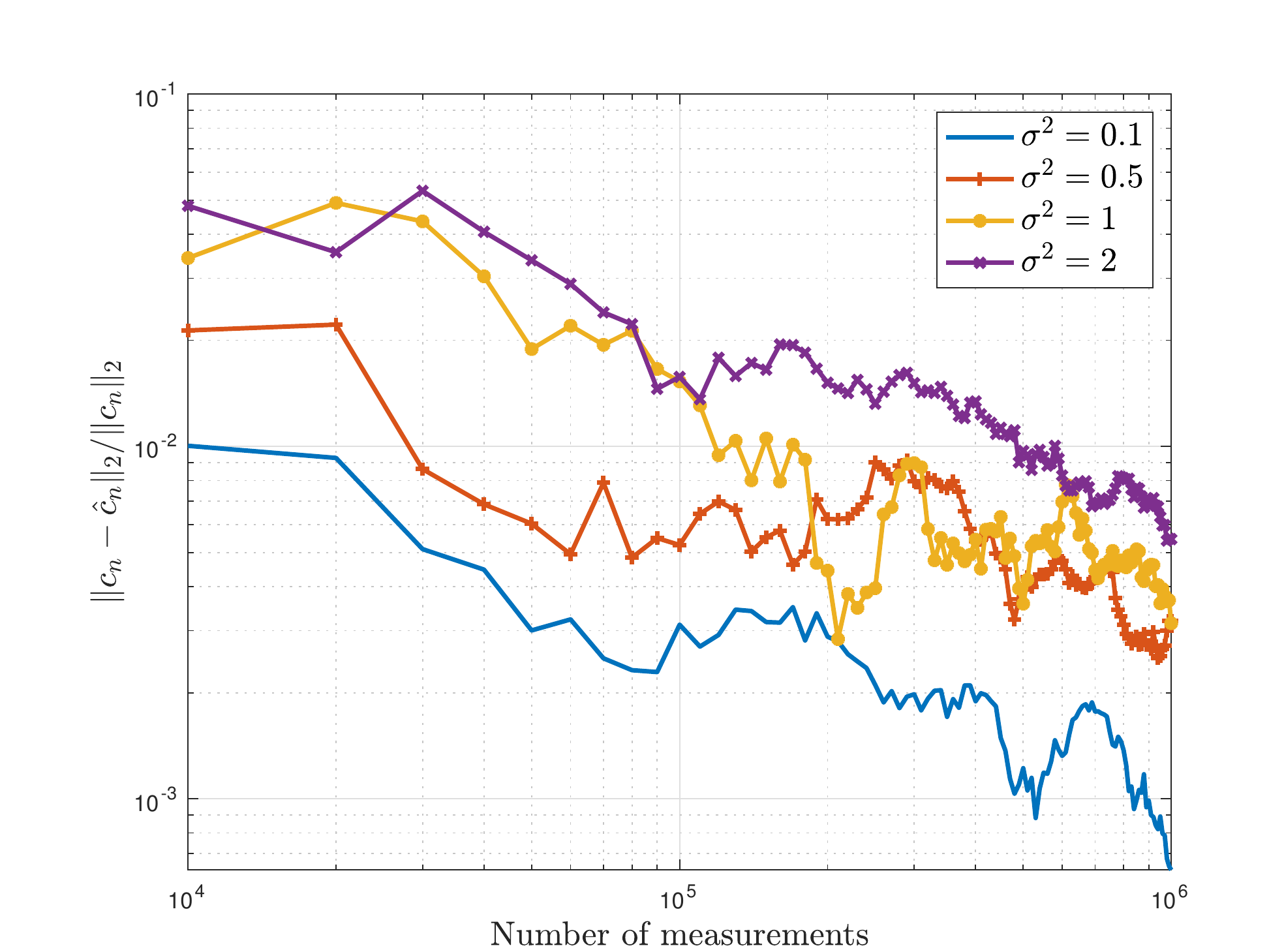}
		\caption{Estimation error of SARX system coefficients}\label{fig:comp2}
	\end{center}
\end{figure}

The estimation error, $\|\cb_n-\hat{\cb}_n\|_2/\|\cb_n\|_2$, as a function of number of measurements, $N$, is depicted in Figure \ref{fig:comp2} for different values of noise variance. As it can be seen from Figure \ref{fig:comp2}, the error again decreases as the number of measurements increases. 
For all values of noise variance, error will eventually decrease  and the estimated values of coefficients converge to the true values.

For the above mentioned example, Table \ref{tab:table4} shows the values of subsystems coefficients for different experiments using  different  values of noise variance. The value of coefficients are very close to the true values, even when the noise variance is high with noise magnitude in average around  76\% of the signal magnitude.

\begin{table}[ht]
	\caption{Identifying submodels' coefficients for different values of noise variance in \textbf{SARX} systems.}
	\label{tab:table4}
	\centering
	\begin{tabular}{c|c|c|c|c|c} 
		\hline \hline      
		\text{submodels' } &true&  variance & variance &variance  &variance \\
		\text{coefficients} & \text{values}& $\sigma^2=0.1$& $\sigma^2=0.5$& $\sigma^2=1$& $\sigma^2=2$\\
		\hline \hline    
		$a_{1}$	&   0.3  & 0.3001 & 0.3011     &  0.2782    &    0.2987  \\
		\hline
		$b_{1}$	&1	&  1.0006     &  1.0022    &   0.9724   &  0.9978     \\
		\hline
		$a_{2}$	&-0.5	&  -0.5008 &   -0.5004 &   -0.4959& -0.5100 \\
		\hline
		$b_{2}$	&-1 &  -0.9995  &  -1.0007  & -1.0012&-1.0096 \\
		\hline \hline
	\end{tabular}
\end{table}

The estimation of noise variance based on the structure of matrix $\Mkpro $ is shown in Table \ref{tab:table3} as $\widehat{\sigma}^2$. As it can be seen from the results obtained, we can efficiently and simultaneously estimate the system's coefficients and the noise variance. 
This illustrates the capability of the proposed algorithm to estimate both system and noise parameters even for large values of noise.

\begin{figure}[h!]
	\begin{center}
		\subfigure[Case 1: true noise variance $\sigma^2= 1$.]{
			\centering\includegraphics[trim={1cm 0.2cm 1cm 1cm},clip,width=0.4\columnwidth]{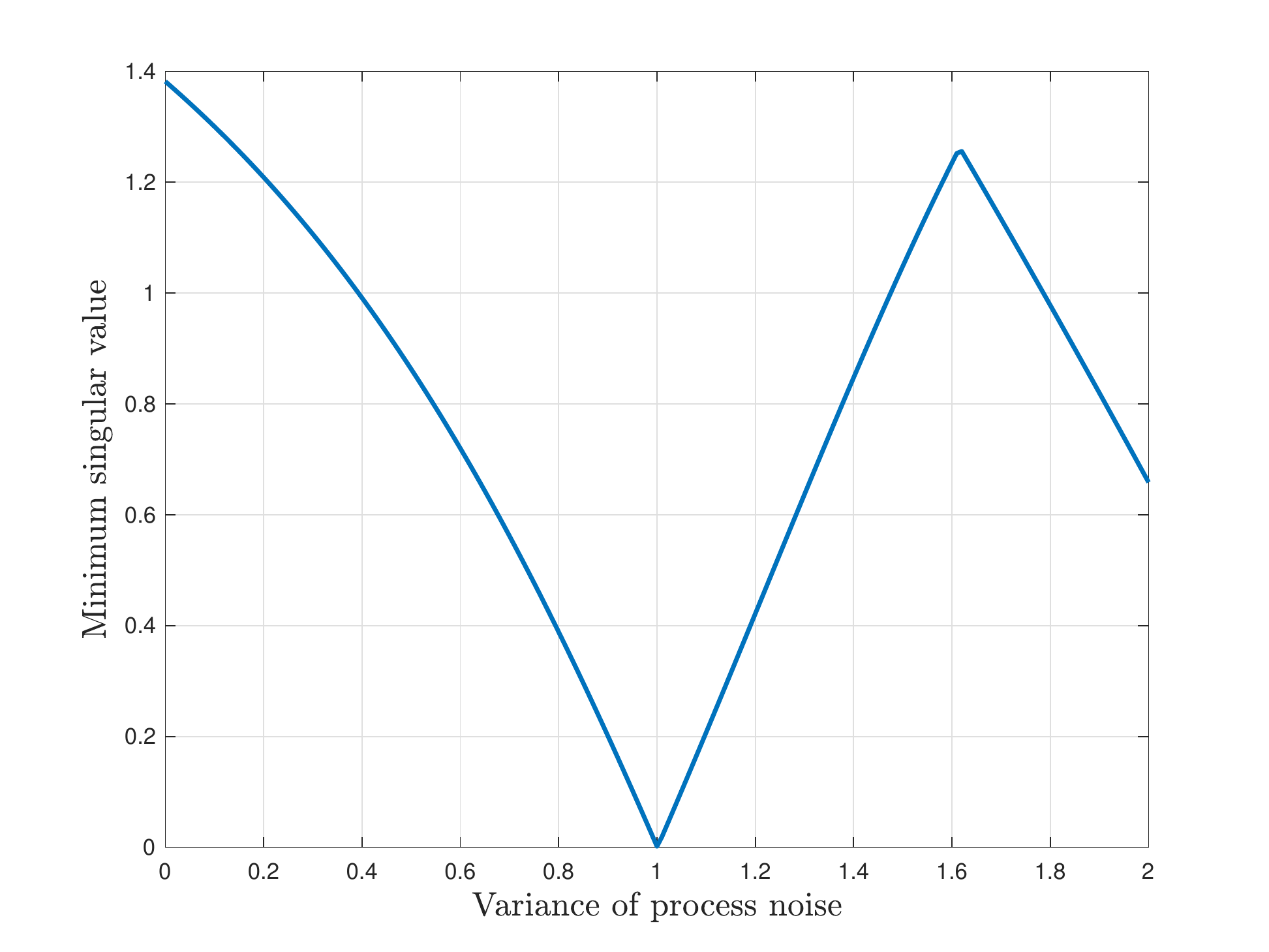} \label{fig:4a}
		}
		\subfigure[Case 2: true noise variance = $\sigma^2=2$.]{
			\centering\includegraphics[trim={1cm 0.3cm 1cm 1cm},clip,width=0.4\columnwidth]{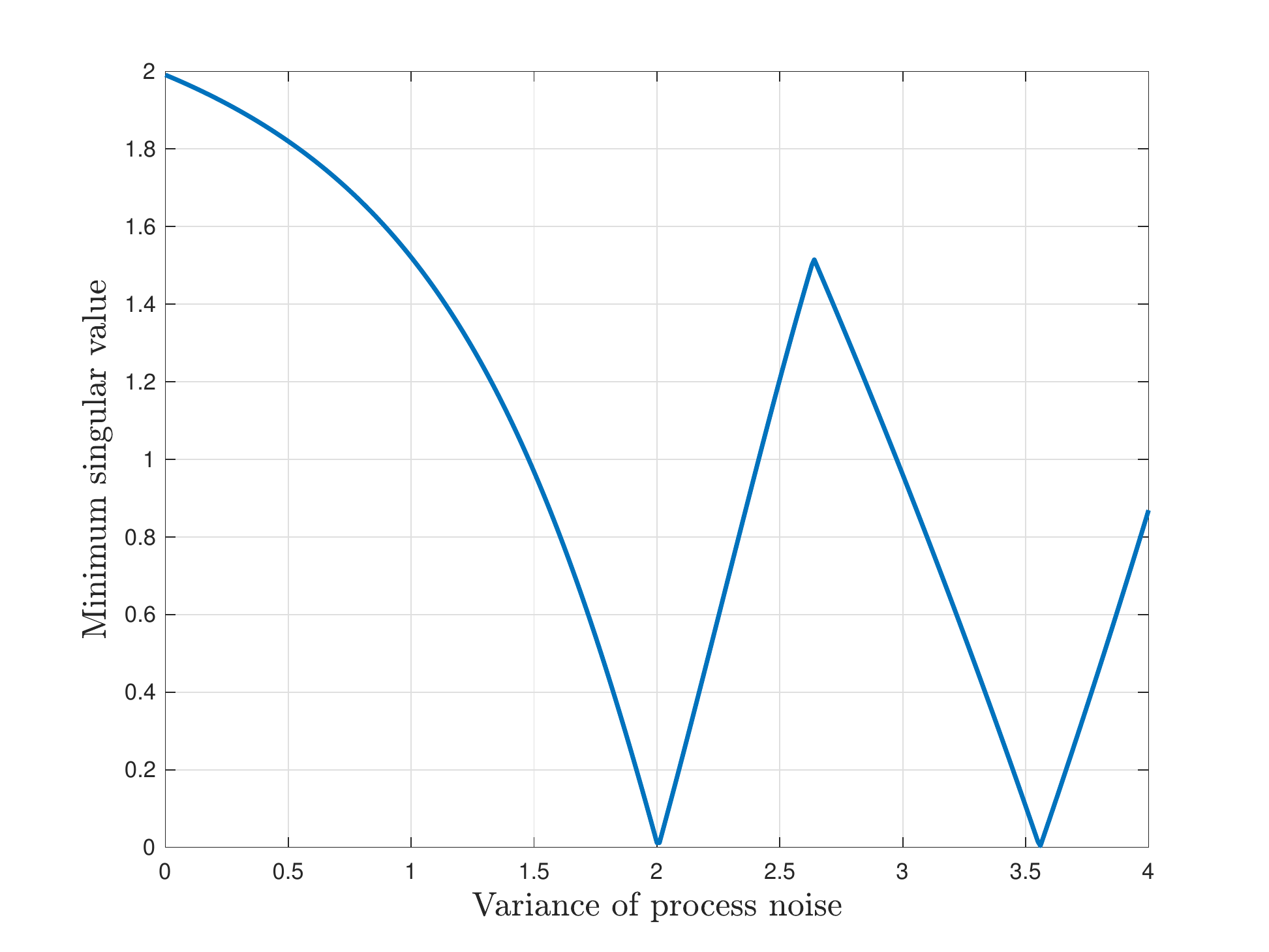} \label{fig:4b}
		}
		\caption{Estimation of noise variance}\label{fig:4}
	\end{center}
\end{figure}

Two examples of the process of estimating the unknown variance of process noise are shown in Fig.~\ref{fig:4}; where Fig.~\ref{fig:4a} is for the case of given data  with process noise of variance 1, and Fig.~\ref{fig:4b} is for data with process noise of variance 2. By taking $\sigma^*$ as the smallest local minimum, the estimated variance for both cases in Fig.~\ref{fig:4a} and Fig.~\ref{fig:4b} is the true values.

\subsubsection{Average Behavior of the Algorithm }
We examine the average behavior of proposed algorithm in this paper for randomly generated stable discrete ARX systems. Systems are randomly generated using the "drss" command of MATLAB, which ensures system poles are random and stable with  possible exception of poles at $1$. Randomly selected system are considered to be of the form of equation \eqref{eq:ex2} with order~1, i.e. $n_a = 1$ and $n_b = 1$, and switched system considered to include two submodels, i.e. $n=2$. 
The average behavior of system is tested for different values of noise variance. In each case, 100 random experiments were run for the total number of measurements~$N = 10^6$. 

The average behavior of the system is shown in Table \ref{tab:table5}. Normalized error is shown by $\beta$ and computed as
\[ \beta = \dfrac{ \norm{\cb_n \, - \, \hat{\cb}_n}} {\norm{\cb_n}}  \]
For different values of variance of noise $\sigma^2 \,=\, 0.1,\, 0.3,\, 0.5,\, 0.7$, the average mean and variance of normalized error are computed and shown in Table \ref{tab:table5}.
In each experiment consisting of 100 run of system, the average of noise to output ratio ($\gamma$)  is computed and shown in Table \ref{tab:table5}.  Note that for some of the randomly generated systems the value of noise to output ratio is close to $1$.
Also average of elapsed time for running the algorithm is shown in Table \ref{tab:table5}.

As we see in this table,  for different values of noise variance the average of difference in identified coefficients in comparison to the true values is really small.  This happens even in the case of large noise to output ratio. For example, in the case of $\sigma^2 \, = \, 0.3$, the average of normalized error is just $0.73\%$ and this is with  having approximately $48\%$ noise to output ratio in average. So, the algorithm can recover the original system efficiently, with very low estimation error and in a short period of time.


\begin{table}
	\caption{Average behavior of Algorithm for different values of noise variance for randomly generated 100 SARX systems.}
	\label{tab:table5}
	\centering
	\begin{tabular}{c|c|c|c|c} 
		\hline \hline      
		\text{Noise variance} &Mean of $\beta$ &  variance of $\beta$ & Mean of $\gamma$  & \text{Mean of elapsed time} \\
		\hline \hline    
		0.1	&   0.0025 & $1.3564e-05$   &0.4259 &  2.9393    \\
		\hline
		0.3	& 0.0073 & $3.3339e-04$ & 0.4799 &  2.8286    \\
		\hline
		0.5	& 0.0083	&  $2.8769e-04$ &  0.5373  &  2.7683  \\
		\hline
		0.7	& 0.0111 &  $5.1764e-04$  & 0.5452   & 2.8888 \\
		\hline \hline
	\end{tabular}
\end{table}

\section{Concluding Remarks} \label{conclusion} 
In this paper we propose a methodology to identify the coefficients of switched autoregressive  and autoregressive exogenous processes and  unknown noise parameters, starting from partial information of the noise and 
given input-output data.  The approach is shown to be particularly efficient in the case of large amount of data. The approach only requires the computation of singular value decomposition of a specially constructed input-output Veronese matrix. The ensuing singular vector is then related to the switched system parameters to be identified.
We prove that the estimated parameters converge to the true ones as the number of measurements grows.
Numerical simulations show a low estimation error, even in the case of large measurement and process noise.
Also, in cases that noise distribution  is not completely known, simulation results show very efficient estimation of unknown noise parameters. 
\bibliographystyle{plain}
\bibliography{Journal_ar_x}

%
\section{APPENDIX } \label{app}

\subsection{Proof of Theorem~\ref{thm1}}
For simplicity of presentation, let
\[
\widehat{M}_{k} \doteq M[\mon (y_k,\ldots,y_{k-n_a},u_{k-1},\ldots,u_{k-n_b})]
\] \vskip 0.1in
We first note that, given the assumptions made on the noise, $u_k$ and $x_k$, the entries of $\widehat{M}_{k}$ have a variance uniformly bounded for all $k$. Moreover
\[
k>l+n_a \Rightarrow \widehat{M}_{k} \text{ and } \widehat{M}_{l} \text{ are independent.}
\] \vskip 0.1in
Hence, by Kolmogorov's Strong  Law of Large Numbers \cite{Sen1993} we have 
\[
\frac{1}{L} \sum_{l=1}^L \widehat{M}_{k+l(n_a+1)} - 	\frac{1}{L} \sum_{l=1}^L E[\widehat{M}_{k+l(n_a+1)}] \longrightarrow 0 ~~\text{ a.s.}
\]
as $L \longrightarrow \infty$. Since 
\[
E[\widehat{M}_{k}] = {M}_{k}~~ \text{ for all positive integer } k
\]
\vskip 0.1in
and applying the results above for $k=1,2,\ldots,n_a+1$, we conclude that 
\[
\frac{1}{N} \sum_{j=1}^N \widehat{M}_{j} - 	\frac{1}{N} \sum_{j=1}^N M_{j} \longrightarrow 0 ~~\text{ a.s.}
\]
as $N \longrightarrow \infty$.
\vskip 0.1in

	
		

\subsection{Convergence Properties of Sums of Dependent Random Variables}
\begin{thm} \label{thmcov} \cite{hu2008convergence}
Let $\{X_n~, n \geq 1\}$ be a sequence of square-integrable random variables and suppose that there exists a
sequence of constants $\{\rho_k~, k \geq 1\}$ such that
\begin{equation} \label{rho}
\sup_{n \geq 1}{|\mathrm{Cov} \,(X_n, \, X_{n+k})|} \leq \rho_k~~~~ k \geq 1
\end{equation}
\vskip 0.1in \noindent
holds. 
Let $\{b_n~, n \geq 1\}$ be a sequence of positive constants satisfying
\[ n=O(b_n) \]
Suppose that
\begin{equation} \label{varvar}
\sum_{n=1}^{\infty}{\dfrac{(\var\,X_n)(\log \, n)^2}{b_n^2}} ~ < ~ \infty
\end{equation}
and
\begin{equation} \label{rhorho}
\sum_{k=1}^{\infty}{\dfrac{\rho_k}{k^q}} ~ < ~ \infty ~~~~~~~~ \text{for some} ~~0 \leq q < 1
\end{equation}
Then 
\begin{equation}
~~~~~~\sum_{i=1}^{n}{\dfrac{X_i - E[X_i]}{b_i}} ~~~~~\text{converges a.s. as}~~~n \to \infty
\end{equation}
\vskip 0.1in \noindent
and if $b_n \uparrow$, the strong law of large number holds, i.e. 
\vskip 0.1in \noindent
\begin{equation}
\lim_{n \to \infty}~
\dfrac{\sum_{i=1}^{n}(X_i - E[X_i])}{b_n} = 0 ~~~~~\text{a.s.}
\end{equation}
\end{thm}

\subsection{Proof of Corollary~\ref{thmwork}}
If assumptions of Corollary \ref{thmwork} hold, the SARX system behaves like linear time varying (LTV) system. 
In general the impulse response of  the discrete linear time varying system at time $k$ is described by 
\begin{equation} \label{ltv}
y_k = \sum_{m=0}^{k}{g(k,m)\, \epsilon_m}~+ R_k(u) ~+ R_k(ic)
\end{equation}    
where $R_k(u)$ is the response to the system input $u$ and $R_k(ic)$ is the  response to initial condition. 
Since the SARX system is uniformly exponentially stable and moments of input and noise are bounded, the responses $R_k(u)$ and $R_k(ic)$ are bounded. On the other hand, the computation of  expected value of output monomials  is a linear combination of the expected values of three responses above. Since the the responses $R_k(u)$ and $R_k(ic)$ are bounded,   in the following reasoning, we  concentrate on the response to noise. So, consider the impulse response of SARX system to be of the form
\begin{equation} \label{ltv1}
y_k = \sum_{m=0}^{k}{g(k,m)\, \epsilon_m}
\end{equation} 
\vskip 0.1in

The discrete time LTV system introduced in equation \eqref{ltv1} is exponentially stable if and only if there exists a constant $M$ and $0 < a < 1$ such that
\begin{equation} \label{sta}
    |g(k,m)| \leq M\, a^{(k-m)}  \quad ~~~ \forall k \geq m .
\end{equation}
\vskip 0.1in \noindent
If we consider the vector matrix format of equation \eqref{ltv1}, i.e. 

 \[
 	\begin{pmatrix}
	y_0  \\
	y_1 \\
	\vdots   \\
	y_N \\
	\end{pmatrix}
	=
	\begin{pmatrix}
	g(0,0) & 0  & \cdots  &  0  \\
	g(1,0)  &        g(1,1)   & \cdots &  0 \\
	\cdots    & \cdots   &\cdots    & \cdots  \\
	g(N,0) &  g(N,1)    &    \cdots      & g(N,N) \\
	\end{pmatrix}
	\begin{pmatrix}
	\epsilon_0  \\
	\epsilon_1 \\
	\vdots   \\
	\epsilon_N \\
	\end{pmatrix}
	\]
or equivalently
\begin{equation} \label{ltv2}
 \yb = A \, \epsilonb
\end{equation} 
\vskip 0.1in \noindent
where $\yb$ is the vector of output measurement and $\epsilonb$ is the vector of noise measurement for all the time. Note that covariance of $\yb$ is computed as 
\begin{equation} \label{cv}
    \mathrm{Cov}(\yb) \, = \, E\,[\yb \, \yb^\top]\, - \, E\,[\yb]\,\, E\,[\yb^\top]\,
\end{equation}
Since we consider noise to have zero mean Normal distribution, output has also Normal distribution and is mean is zero. So, the covariance of  $\yb$ in equation \eqref{cv} is 
\begin{equation} \label{cv1}
    \mathrm{Cov}(\yb) \, = \, E\,[\yb \, \yb^\top]\, = A \, E\,[\epsilonb \, \epsilonb^\top]\, A^\top = m_2 \, A \, A^\top 
\end{equation}
where $m_2$ is the second moment or variance of the noise. 

\vskip 0.1in
Considering the case where assumptions of Theorem \ref{thmwork} hold, we use several steps of reasoning to show the conditions of Theorem \ref{thm11} are satisfied and  that the algorithm in this paper converges.
\begin{enumerate}
	\item 
	We assumed that SARX system is uniformly exponentially stable, input is bounded and moments of noise up to order $4n$ are bounded. 
	\item Step (1) leads to having 
	the expected value of output monomial up to order $2n$ bounded and therefore the variance of output monomials is bounded.
	\item Step (1) and (2) lead to the conditions of Theorem \ref{thm11} being satisfied.
	%
	%
%
	\item As a result, the strong  law of large numbers holds for 
	the monomials of system output up to order $2n$ and  the average of the results obtained from a large number of experiments converges to the desired value in equation \eqref{eq:Mk2} almost surely. 
In other words, Theorem \ref{thmwork} holds and  
$~ \widehat\MNpro - \MNpro  \longrightarrow ~0~~ \text{ a.s.}$, as  $ N\rightarrow \infty$.
\end{enumerate}

Now, we prove every step from reasoning above:
\vskip 0.1in  \noindent
\textbf{1. Input and moments of noise are bounded:}

Based on Assumption \ref{ass2}, input is given and bounded, and moments of noise up to order $4n$ are bounded. 
\vskip 0.1in \noindent
\textbf{2. Expected value of  output monomial up to order $2n$ are bounded:}

Switched system is uniformly exponentially stable and input is bounded. Moreover, noise is assumed to have zero mean  Normal distribution with bounded moments, so output moments are also bounded. 
%
Therefore, the output monomial $z_k$ as in equation \eqref{mon}, is a monomial of Normal random variables, so its expectation is bounded. 
%
As a result, the variance of output monomials is also bounded.

\vskip 0.1in \noindent
\textbf{3. Conditions of Theorem \ref{thm11} are satisfied.}

Lets consider $z_k$ as a monomial of output up to order $2n$ as defined in equation \eqref{mon}, then
\begin{equation} \label{covi}
    \mathrm{Cov} \,(z_k, \, z_{k+l})\, = \, E\,[z_k \, z_{k+l}]\, - \, E\,[z_k]\, E\,[ z_{k+l}]
\end{equation}
\vskip 0.1in \noindent
Since the noise is considered to be zero mean Normal, output monomials have multivariate normal distribution. So,  we are able to compute higher order moments of the multivariate normal distribution in terms of its covariance matrix based on Isserlis' theorem \cite{isserlis1918formula} which is as follows:
\vskip 0.1in
\textbf{Isserlis' Theorem} \cite{isserlis1918formula}
If $(X_1, X_2, \cdots, X_{2n+1}),~\forall n=1, 2, \cdots$ are zero mean multivariate
Normal random variables, then
\begin{equation}
   E\,[ X_1 \, X_2 \, \cdots \, X_{2n}] \, = \, \sum \, \prod E\,[ X_i X_j]
\end{equation}
and
\begin{equation}
   E\,[ X_1 \, X_2 \, \cdots \, X_{2n+1}] \, = \, 0
\end{equation}
\noindent
where the notation $\sum \, \prod$ means summing over all distinct ways of partitioning $X_1, X_2, \cdots, X_{2n}$ into pairs $ X_i, X_j$, which yields to $(2n)!/(2^n n!)$ terms in the sum.

\vskip 0.1in
By using the results of Isserlis Theorem for computing the value $E\,[z_k \, z_{k+l}]\,$ in equation \eqref{covi}, we have
\begin{equation} \label{cov2}
     \mathrm{Cov} \,(z_k \, z_{k+l}) \, = \, \sum_{h=1}^{w}{q_h \sigma_{i_h\, j_h}}
\end{equation}
where $|i_h\,- j_h|\, \geq \, l$ and $w \, \leq (4n)!/(2^{2n}\, (2n)!)$ considering that the maximum order of monomial $z_k$ and $z_{k+l}$ can each be $2n$. 
Distance $|i_h\,- j_h|$  is the distance from the diagonal of covariance matrix of output, which is introduced by equation \eqref{cv1}, and
$\sigma_{i\, j}$ is the ${ij}^\text{th}$ entry  of the covariance matrix of output. 
Note that $\sigma_{i_h\, j_h}$ is the part of $\sigma_{i\, j}$
which $|i_h\,- j_h|\, \geq \, l$, and 
$q_h$ is  the remaining part.
So  in equation \eqref{cov2} we  consider the elements of covariance matrix of $y$ with largest distance as $\sigma_{i_h\, j_h}$, and put the rest as $q_h$.
\vskip 0.05in

Since the system is uniformly exponentially stable, the system impulse response decays exponentially, therefore by going farther from diagonals of the covariance matrix the entries of covariance matrix of output $\sigma_{i\, j}$s decrease exponentially
and distance $|i_h\,- j_h|$  decays proportionally with the distance from the diagonal.

First we prove that for $ Cov \,(z_k, \, z_{k+l}) \, = \, \sum_{h=1}^{w}{q_h \sigma_{i_h\, j_h}}$  in equation \eqref{cov2}, we always have  $|i_h\,- j_h|\, \geq \, l$.
For computing  $E\,[z_k \, z_{k+l}]\,$ in equation \eqref{covi} there are two cases that might happen:
\begin{enumerate}
    \item The case that time indices of $\sigma_{i_h\, j_h}$ involved in computing $E\,[z_k \, z_{k+l}]\,$, are always with the interval $|i_h\,- j_h|\, \geq \, l$ 
    \item The case that time some indices of $\sigma_{i_h\, j_h}$ involved in computing $E\,[z_k \, z_{k+l}]\,$, are  in the interval $|i_h\,- j_h|\, < \, l$ 
\end{enumerate}
First case lines with the fact that in computing the expected value of each pair based on Isserlis' theorem, there exists at least one entry of covariance matrix called $\sigma_{i_h\, j_h}$ that the distance $|i_h\,- j_h|\, \geq \, l$. For the second case, if there is no entry with the  distance $|i_h\,- j_h|\, \geq \, l$, then that means the entry is separated into the multiplication of terms. In other words, there is one term that is related to the first monomial $z_k$, and the other term is related to the second monomial $z_{k+l}$. So that the multiplication of these terms is cancelled by the $E\,[z_k]\, E\,[ z_{k+l}]$ term in equation \eqref{covi}.
	
	\vskip 0.05in
	
Now that we have proved there is always distance $|i_h\,- j_h|\, \geq \, l$	in computing the   $ Cov \,(z_k, \, z_{k+l}) \,$  in equation \eqref{cov2}, it is time to find an upper bound for the   $ Cov \,(z_k, \, z_{k+l}) \,$  and call it $\rho_l$.
\vskip 0.1in
As we have shown in equation \eqref{sta}, $    |g(k,m)| \leq M\, a^{(k-m)}  ~~~ \forall k \geq m $, where  $M$ is a constant and $0 < a < 1$. 
Because the distance $|i_h\,- j_h|\, \geq \, l$, then 
\[\sigma_{i_h\, j_h}  \leq \tilde{M} \, a^l \]
for some constant $\tilde{M}$.
So,

\begin{equation} \label{cov3}
     \mathrm{Cov} \,(z_k, \, z_{k+l}) \, = \, \sum_{h=1}^{w}{q_h \sigma_{i_h\, j_h}} \, \leq  C\, a^l
\end{equation}
where $C$ is a constant. Therefore we pick 
\begin{equation} \label{rhoo}
\rho_l(C, a) = C\, a^l
\end{equation}
where 
$0 < a < 1$. 
\vskip 0.1in \noindent
Now, we prove 
\begin{equation} \label{pf1}
\sum_{k=1}^{\infty}{\dfrac{(\var\,z_k)(\log \, k)^2}{k^2}} ~ < ~ \infty
\end{equation}
\vskip 0.1in \noindent
holds. 
In Step (2) we have proved that variance of output monomials ($\var\,z_k$) is bounded. Moreover, $ \sum_{k=1}^{\infty} \dfrac{(\log \, k)^2}{k^2}$
is known to be bounded and converges, so equation \eqref{pf1} holds.

The last condition of Theorem \ref{thm11} that we need to prove is \begin{equation} \label{pf2}
    \sum_{l=1}^{\infty}{\dfrac{\rho_l}{l^q}} ~ < ~ \infty ~~~~~~~~ \text{for some} ~~0 \leq q < 1
\end{equation}

\noindent
By considering $\rho_l= C\, a^l$ and $q=0$, the equation \eqref{pf2} becomes 
 
\begin{equation} \label{rhorho1}
\sum_{l=1}^{\infty}{\dfrac{\rho_l}{l^q}} ~ = ~ \sum_{l=1}^{\infty}{\rho_l}~ = ~ \sum_{l=1}^{\infty}{C\, a^l} ~ < ~\infty ~~~~~~~~  ~~0 \leq a < 1
\end{equation}
\vskip 0.1in \noindent
So, we have shown that equation \eqref{pf2} holds.

\vskip 0.1in \noindent
	\textbf{4. Strong  law of large numbers holds for system output monomials} 
	
	As it has shown in previous steps, the conditions of Theorem \ref{thmcov} are satisfied for SARX system identification in this paper, so  
	\begin{equation}
\lim_{N \to \infty}~
\dfrac{\sum_{k=1}^{N}(X_k - E[X_k])}{k} = 0 ~~~~~\text{a.s.}
\end{equation}
	In other words, strong law of large numbers holds for the system output and its monomials. 

\vskip 0.1in
Now that we have proved strong law of large numbers holds for  monomials of system output, the direct result is that strong law of large numbers holds for the $M[\mon (y_k,\ldots,y_{k-n_a},u_{k-1},\ldots,u_{k-n_b})]$ and accordingly as  $ N\rightarrow \infty$,
\[
 \frac{1}{N} \sum_{k=1}^N M[\mon (y_k,\ldots,y_{k-n_a},u_{k-1},\ldots,u_{k-n_b})]-  \frac{1}{N} \sum_{k=1}^N E[M[\mon (y_k,\ldots,y_{k-n_a},u_{k-1},\ldots,u_{k-n_b})]]  \longrightarrow ~0~~ \text{ a.s.}
\] 
which based on notations in Theorem \ref{thmwork}, Lemma \ref{lemma:Mk2} and equation \eqref{eq:Mk2} it is: as  $ N\rightarrow \infty$,
\[
	\widehat\MNpro - \MNpro  \longrightarrow ~0~~ \text{ a.s.}  
\]

\clearpage
\section*{Author Biography}
%
%

\begin{biography}
{\includegraphics[width=66pt,height=86pt]{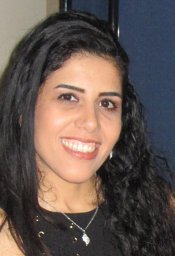}}{\textbf{Sarah Hojjatinia.}
Sarah Hojjatinia is currently a Ph.D. candidate in electrical engineering at the Pennsylvania State University, University Park, PA. 
She received the B.S. and M.S. degrees in electrical engineering from K. N. Toosi University of Technology, Tehran in 2009 and 2013 respectively, She started her Ph.D. program in electrical engineering with dual M.S. degree in mechanical engineering at the Pennsylvania State University, University Park, PA in 2015, where she received the M.S. degree in mechanical engineering in 2018. 
Her research interests include system identification, machine learning, control and optimization with applications in behavioral sciences and data analysis. 
She is currently a member of the IEEE Control Systems Society, IEEE Women in Engineering and IEEE Young Professionals.
} 
\end{biography}

\begin{biography}
{\includegraphics[width=66pt,height=86pt]{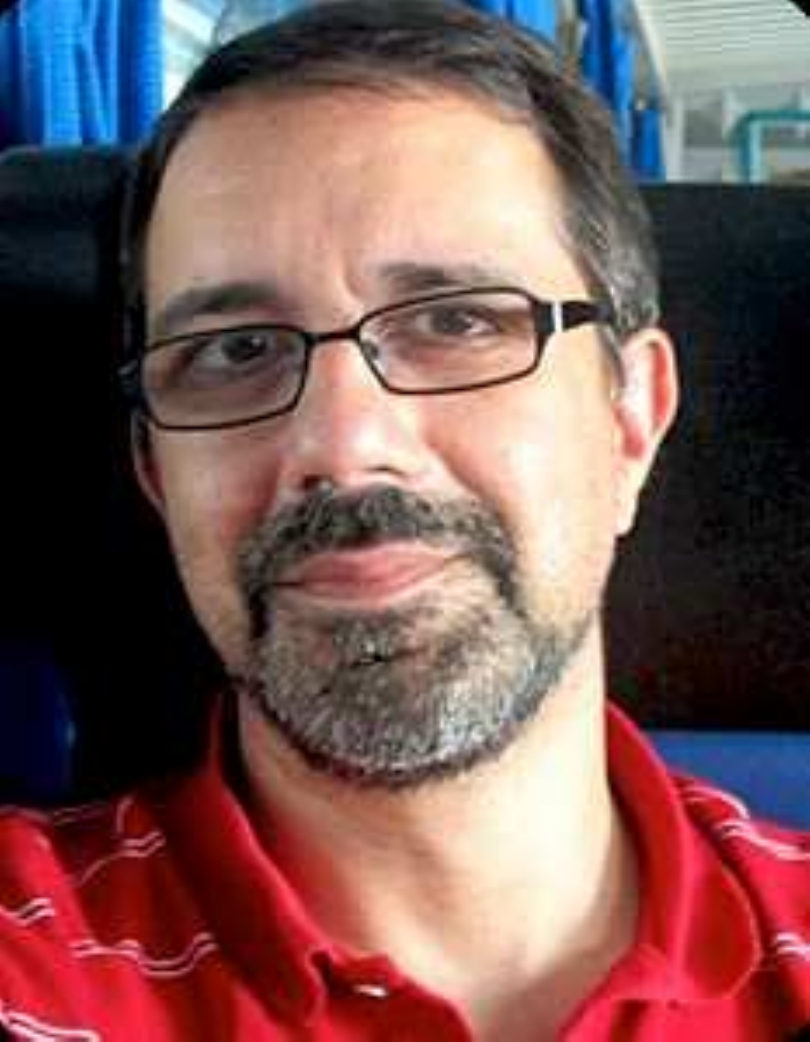}}{\textbf{Constantino Lagoa.}
Constantino M. Lagoa received the B.S. and M.S. degrees from the
Instituto Superior Tecnico, Tech- ical University of Lisbon, Portugal
in 1991 and 1994, respectively, and the Ph.D. degree from the University
of Wisconsin at Madison in 1998. He joined the Electrical Engineering
Department of Pennsylvania State University, University Park, PA, in
August 1998, where he currently holds the position of Professor. He
has a wide range of research interests including robust optimization and
control, chance constrained optimization, controller design under risk
specifications, system identification and control of computer networks. Dr. Lagoa has served as Associate
Editor of IEEE Transactions on Automatic Control (2012-2017) and IEEE Transactions on Control systems Technology (2009-2013) and he is currently Associate Editor of Automatica.}
\end{biography}

\begin{biography}
{\includegraphics[width=66pt,height=86pt]{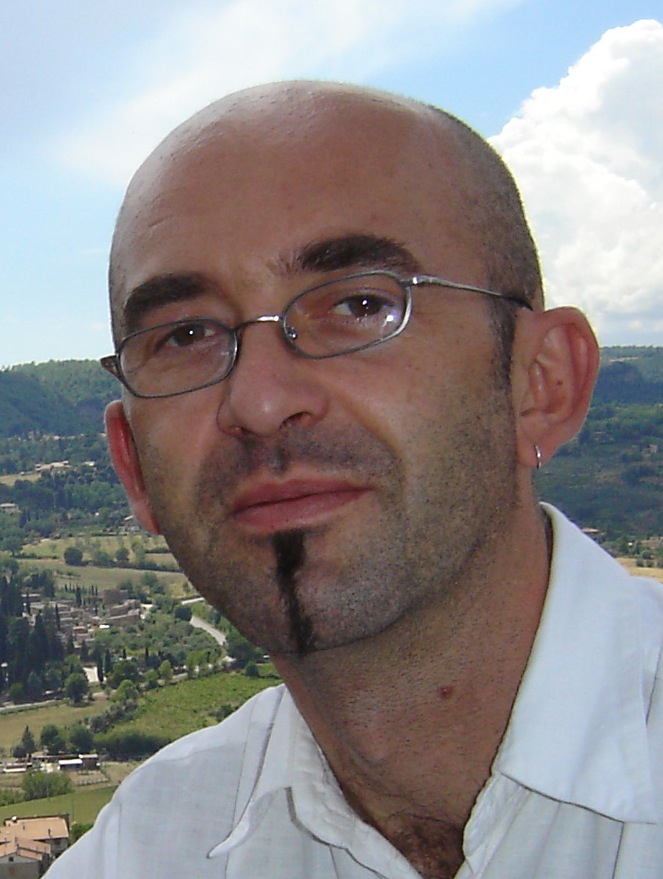}}{\textbf{Fabrizio Dabbene.}
Fabrizio Dabbene received the Laurea degree in 1995 and the Ph.D. degree
in 1999, both from Politecnico di Torino, Italy. He is currently Senior
Researcher at the CNR-IEIIT institute. His research interests include
randomized and robust methods for systems and control, and modeling of
environmen- tal systems. He published more than 100 research papers and
two books, and is recipient of the 2010 EurAgeng Outstanding Paper
Award. He served as Associate Editor for Automatica (2008-2014) and IEEE
Transactions on Automatic Control (2008- 2012). Dr. Dabbene is a Senior
Member of the IEEE, and has taken various responsibilities within the
IEEE-CSS: he served as elected member of the Board of Governors
(2014-2016) and as Vice President for Publications (2015- 2016).
}
\end{biography}

\end{document}